\documentclass[%
reprint,
superscriptaddress,
showpacs,preprintnumbers,
 amsmath,amssymb,
 aps,
prd,
]{revtex4-1}

\usepackage{graphicx}
\usepackage{dcolumn}
\usepackage{bm}
\usepackage{bbold}
\usepackage{amssymb,amsmath}
\usepackage{hyperref}

\usepackage{color}
\usepackage{amsfonts}
\usepackage{subfigure}
\usepackage{array}

\newcommand{\Tr}{\ensuremath{\operatorname{Tr}}}

\newcolumntype{L}{>{\centering\arraybackslash}m{3cm}}

\definecolor{bjcol}{rgb}{1,.44,0.13}

\definecolor{blue}{rgb}{0,0,1}

\definecolor{green}{rgb}{0,1,0}

\definecolor{red}{rgb}{1,0,0}

\definecolor{gray}{rgb}{.5,.5,.5}

\definecolor{darkgreen}{rgb}{.0,.5,.0}

\def\Fig#1{Fig.~\ref{#1}}

\def\Eq#1{Eq.~(\ref{#1})}

\def\eqref#1{(\ref{#1})}

\def\tab#1{Tab.~\ref{#1}}

\def\sec#1{Sec.~\ref{#1}}
\def\app#1{Appendix~\ref{#1}}

\def\lA0{{\langle A_0 \rangle}}
\def\bA0{{\bar{A}_0}}

\def\0#1#2{\frac{#1}{#2}}

\graphicspath{{./figures/}{./}}

\begin{document}

\preprint{}

\title{Critical behaviors of the $O(4)$ and $Z(2)$ symmetries in the QCD phase diagram}

\author{Yong-rui Chen}
\affiliation{School of Physics, Dalian University of Technology, Dalian, 116024,
  P.R. China}

\author{Rui Wen}
\affiliation{School of Physics, Dalian University of Technology, Dalian, 116024,
  P.R. China}

\author{Wei-jie Fu}
\email{wjfu@dlut.edu.cn}
\affiliation{School of Physics, Dalian University of Technology, Dalian, 116024,
  P.R. China}


\begin{abstract}

In this work we have studied the QCD phase structure and critical dynamics related to the 3-$d$ $O(4)$ and $Z(2)$ symmetry universality classes in the two-flavor quark-meson low energy effective theory within the functional renormalization group approach. We have employed the expansion of Chebyshev polynomials to solve the flow equation for the order-parameter potential. The chiral phase transition line of $O(4)$ symmetry in the chiral limit, and the $Z(2)$ line of critical end points related to the explicit chiral symmetry breaking are depicted in the phase diagram. Various critical exponents related to the order parameter, chiral susceptibilities and correlation lengths have been calculated for the 3-$d$ $O(4)$ and $Z(2)$ universality classes in the phase diagram, respectively. We find that the critical exponents obtained in the computation, where a field-dependent mesonic nontrivial dispersion relation is taken into account, are in quantitative agreement with results from other approaches, e.g., the conformal bootstrap, Monte Carlo simulations and $d=3$ perturbation expansion, etc. Moreover, the size of the critical regime in the QCD phase diagram is found to be very small.

\end{abstract}

\maketitle


\section{Introduction}
\label{sec:intro}

Significant progress has been made in studies of QCD phase structure over the last decade, both from the experimental and theoretical sides; see, e.g. \cite{Stephanov:2007fk,Friman:2011zz,Luo:2017faz,Andronic:2017pug,Fischer:2018sdj,Bzdak:2019pkr,Fu:2019hdw,Bazavov:2020bjn,Borsanyi:2020fev,Fu:2021oaw}. One of the most prominent features of the QCD phase structure is the probable presence of a second order critical end point (CEP) in the phase diagram spanned by the temperature $T$ and baryon chemical potential $\mu_B$ or densities, which separates the first order phase transition at high $\mu_B$ from the continuous crossover at low $\mu_B$ \cite{Stephanov:2007fk}. The existence and location of CEP are, however, still open questions, whose answers would definitely help us to unravel the most mysterious veil related to the properties of strongly interacting matter under extreme conditions. The Beam Energy Scan (BES) Program at the Relativistic Heavy Ion Collider (RHIC) is aimed at searching for and locating the critical end point, where fluctuation observables sensitive to the critical dynamics, e.g., high-order cumulants of net-proton, net-charge, net-kaon multiplicity distributions, have been measured  \cite{Adamczyk:2013dal, Adamczyk:2014fia, Luo:2015ewa, Adamczyk:2017wsl}. Notably, a non-monotonic dependence of the kurtosis of the net-proton multiplicity distribution on the beam energy with $3.1\sigma$ significance in central collisions has been reported by the STAR collaboration recently \cite{Adam:2020unf}. 

On the other hand, lattice QCD simulations have provided us with a plethora of knowledge about the QCD phase structure, e.g., the crossover nature of the chiral phase transition at finite $T$ and vanishing $\mu_B$ with physical current quark mass \cite{Aoki:2006we}, pseudo-critical temperature \cite{Borsanyi:2013bia,Bazavov:2014pvz}, curvature of the phase boundary \cite{Bazavov:2018mes,Borsanyi:2020fev}, etc. Because of the notorious sign problem at finite chemical potential, the reliability regime of lattice calculations is restricted to be $\mu_B/T\lesssim 2\sim3$, where no CEP has been found. Free from the sign problem, the first-principle functional approaches, e.g, the functional renormalization group (fRG) and Dyson-Schwinger equations (DSE), could potentially extend the regime of reliability to $\mu_B/T \sim 4$ \cite{Fischer:2018sdj,Fu:2019hdw}. With benchmark tests of observables at finite $T$ and low $\mu_B$ in comparison to lattice calculations, e.g., the quark condensate, curvature of the phase boundary, etc., functional approaches, both fRG and DSE, have predicted a CEP located in a region of $450\,\mathrm{MeV} \lesssim\mu_B\lesssim 650\,\mathrm{MeV}$ \cite{Fischer:2018sdj,Fu:2019hdw,Isserstedt:2019pgx,Gao:2020qsj,Gao:2020fbl} recently.

An alternative method used to circumvent the possible location of CEP, is to determine the critical temperature $T_c$ of the chiral phase transition in the chiral limit, more specifically, i.e., massless light up and down quarks and a physical strange quark mass. Since it is believed that the value of $T_c$ sets an upper bound for the temperature of CEP \cite{Halasz:1998qr,Buballa:2018hux}. Very recently, the critical temperature $T_c$ in the chiral limit has been investigated and its value is extrapolated from both lattice simulations \cite{Ding:2019prx} and functional approach \cite{Braun:2020ada}. Moreover, further lattice calculations indicate that axial anomaly remains manifested at $T\approx 1.6\,T_c$, which implies that the chiral phase transition of QCD in the chiral limit is of 3-$d$ $O(4)$ universality class \cite{Ding:2020xlj}; see, e.g., \cite{Pisarski:1983ms} for more discussions about the relation between the axial anomaly and the symmetry universality classes.

In this work, we would like to study the QCD phase structure in the chiral limit and finite current quark mass, i.e., with a finite pion mass, in the two-flavor quark-meson low energy effective theory (LEFT) within the fRG approach. For more discussions about the fRG approach, see, e.g., QCD related reviews \cite{Berges:2000ew,Pawlowski:2005xe,Schaefer:2006sr,Gies:2006wv,Rosten:2010vm,Braun:2011pp,Pawlowski:2014aha,Dupuis:2020fhh}. In contrast with the lattice simulation and the first-principle fRG-QCD calculation \cite{Ding:2019prx,Braun:2020ada}, the chiral limit could be accessed strictly in the LEFT. Furthermore, we would also like to study the critical behaviors of the 3-$d$ $O(4)$ and $Z(2)$ universality classes, including various critical exponents, which belong to the second-order chiral phase transitions in the chiral limit and at the critical end point with finite quark mass, respectively. To that end, we expand the effective potential of order parameter as a sum of Chebyshev polynomials in the computation of fRG flow equations; see \cite{AndreasRisch:2013} for more details. The Chebyshev expansion of solutions to a set of integrodifferential equations is, in fact, a specific formalism of more generic pseudo-spectral methods \cite{Boyd:2000}, and see also, e.g., \cite{Borchardt:2015rxa,Borchardt:2016pif,Knorr:2020rpm} for applications of pseudo-spectral methods in the fRG.

In fact, another two numerical methods are more commonly used in solving the flow equation for the effective potential: one is the Taylor expansion of the effective potential around some value \cite{Pawlowski:2014zaa,Yin:2019ebz}, and the other discretization of the effective potential on a grid \cite{Schaefer:2004en}. The (dis)advantages of these two methods are distinct. The former is liable to implementation of the numerical calculations, but short of global properties of the effective potential, that is, however, indispensable to studies of chiral phase transition in the chiral limit or around CEP; the latter is encoded with global information on the potential, but it loses numerical accuracy near the phase transition point which is necessary especially for the computation of critical exponents. The Chebyshev expansion used in this work combines the merits from both approaches, i.e., the global potential and the numerical accuracy, and thus it is very suitable for the studies of critical behaviors in the QCD phase diagram. Remarkably, a discontinuous Galerkin scheme has been applied in the context of fRG recently \cite{Grossi:2019urj}, which is well-suited for studies of the first-order phase transition.

This paper is organized as follows: In \sec{sec:QM} we briefly introduce the flow equations in the quark-meson LEFT and the method of the Chebyshev expansion for the effective potential. The obtained phase diagram and QCD phase structure are presented and discussed in \sec{sec:phasediagram}. In \sec{sec:exponent} scaling analyses for the the 3-$d$ $O(4)$ and $Z(2)$ universality classes are performed, and various critical exponents are obtained. We also discuss the size of the critical regime there. In \sec{sec:summary} we give a summary and conclusion. Some threshold functions and anomalous dimension in the flow equations, and some relations for the Chebyshev polynomials are collected in \app{app:threshold} and \app{app:cheby}, respectively.

\section{Functional renormalization group and the low energy effective theories}
\label{sec:QM}

Thanks to the Wilson's idea of the renormalization group (RG), see, e.g., \cite{Wilson:1973jj}, it has been well known that usually the active degrees of freedom are quite different, when the energy scale of a system evolves from a hierarchy into another. The relevant dynamics in different hierarchies are connected with each other through the evolution of RG equations. To be more specific, in QCD the partonic degrees of freedom, i.e., the quarks and gluons, in the high energy perturbative regime are transformed into the collective hadronic ones in the nonperturbative region of low energy, with the RG scale evolving from the ultraviolet (UV) to infrared (IR) limits \cite{Weinberg:1978kz}, and see also, e.g., \cite{Gies:2001nw,Gies:2002hq,Pawlowski:2005xe,Floerchinger:2009uf,Braun:2014ata,Mitter:2014wpa,Cyrol:2017ewj,Eser:2018jqo,Fu:2019hdw} for recent development of the relevant ideas within the fRG approach. When the momentum or RG scale is below, say $\sim 1$ GeV, which is related to a narrow transition region from the perturbative to nonperturbative QCD, calculated results of Yang-Mills theory and QCD in Landau gauge indicate that the gluons develop a finite mass gap and decouple from the system, and see, e.g. \cite{Mitter:2014wpa,Cyrol:2016tym,Fu:2019hdw,Huber:2020keu} for more details. As a consequence, contributions to the flow equations of effective action from the glue sector could be safely neglected, if the initial evolution scale is set at a UV scale $\Lambda \lesssim 1$ GeV. 

Hence, within the fRG approach, one is left with the flow equation for the low energy effective theory, which reads
\begin{align}
\partial_t\Gamma_k[\Phi]=&-\mathrm{Tr}\Big(G_{q\bar q,k}\partial_t R_{q,k}\Big)+\frac{1}{2}\mathrm{Tr}\Big(G_{\phi\phi,k}\partial_t R_{\phi,k}\Big)\,,\label{eq:flow}
\end{align}
with the RG scale $k$ and the RG time defined as $t=\ln(k/\Lambda)$. Apparently, \Eq{eq:flow} is an ordinary differential equation for the $k$-dependent effective action, $\Gamma_k[\Phi]$, the arguments $\Phi=(q,\bar q,\phi)$ of which are the quark and mesonic fields in the LEFT. The equation in \Eq{eq:flow}, which describes the evolution of the effective action with the RG scale, is also well known as the Wetterich equation \cite{Wetterich:1992yh}, see also \cite{Ellwanger:1993mw,Morris:1993qb}. The flow receives contributions from both the quark and mesonic degrees of freedom, as shown on the r.h.s. of \Eq{eq:flow}, where $G_{q\bar q,k}$ and $G_{\phi\phi,k}$ are the $k$-dependent full quark and meson propagators, respectively, and are related to the quadratic derivatives of $\Gamma_k[\Phi]$ with respect to their respective fields, viz.
\begin{align}
  G_{\phi\phi/q\bar{q}}[\Phi]=\left( \frac{1}{\frac{\delta^2\Gamma_k[\Phi]}{\delta\Phi^2}+R_{\Phi,k}} \right)_{\phi\phi/q\bar{q}}\,. \label{eq:props}
\end{align}
where $R_{q,k}$ and $R_{\phi,k}$ as well as in \Eq{eq:flow} are the IR regulators, which are employed to suppress quantum fluctuations of momenta $q \lesssim k$, and their explicit expressions used in the work are given in Eqs. (\ref{eq:Rphi}) and (\ref{eq:Rq}). Moreover, interested readers could refer to QCD related fRG review articles \cite{Berges:2000ew,Pawlowski:2005xe,Schaefer:2006sr,Gies:2006wv,Rosten:2010vm,Braun:2011pp,Pawlowski:2014aha,Dupuis:2020fhh} for more details about the formalism of fRG, and also \cite{Braun:2007bx,Braun:2008pi,Braun:2009gm,Mitter:2014wpa,Braun:2014ata,Cyrol:2016tym,Cyrol:2017ewj,Cyrol:2017qkl,Fu:2019hdw,Braun:2020ada,Fu:2021oaw} for recent progress on relevant studies.

In this work, we adopt a truncation for the effective action in \Eq{eq:flow} as follows
\begin{align}
\Gamma_k[\Phi]=&\int_x \bigg\{Z_{q,k}\bar{q} \big (\gamma_\mu \partial_\mu -\gamma_0\hat\mu \big )q+\frac{1}{2}Z_{\phi,k}(\rho) \big (\partial_\mu \phi \big)^2 \nonumber\\[2ex]
&+h_{y,k}\bar{q}\big(T^0\sigma+i\gamma_5\vec{T}\cdot \vec{\pi}\big)q+V_k(\rho)-c\sigma \bigg\}\,,\label{eq:action}
\end{align}
with the shorthand notation $\int_{x}=\int_0^{1/T}d x_0 \int d^3 x$, where the quark field $q=(u\,,d)^{T}$ and the meson field $\phi=\left(\sigma,\vec{\pi}\right)$ are in the fundamental and adjoint representations of $SU(N_f)$ in the flavor space with $N_f=2$, respectively. They interact with each other via a Yukawa coupling with a coupling strength $h_{Y,k}$, where the subscript $_Y$ is used to distinguish it from the reduced external field $h$ in \Eq{eq:th}. Here $T^i$ ($i=1\,,2\,,3$) are the generators of $SU(2)$ with $\Tr(T^{i}T^{j})=\frac{1}{2}\delta^{ij}$ and $T^{0}=\frac{1}{\sqrt{2N_{f}}}\mathbb{1}_{N_{f}\times N_{f}}$. Note that both the effective potential $V_k(\rho)$ and the mesonic wave function renormalization $Z_{\phi,k}(\rho)$ in \Eq{eq:action} depend on the meson field by means of $\rho=\phi^2/2$, which are $O(4)$ invariant. $Z_{q,k}$ is the quark wave function renormalization. Notice that the term linear in the order parameter field, i.e., $-c\sigma$ in \Eq{eq:action}, breaks the chiral symmetry explicitly, and thus here $c$ is essentially an external ``magnetic'' field in the language of magnetization. Moreover, $\hat\mu=\mathrm{diag}(\mu_u,\mu_d)$ is the matrix of quark chemical potentials in the flavor space, and $\mu=\mu_u=\mu_d$ is assumed throughout this work, which is related to the baryon chemical potential via $\mu=\mu_B/3$. For more discussions about the quark-meson LEFT in \Eq{eq:action} or its extensions, e.g., Polyakov-loop quark-meson LEFT, QCD assisted LEFT, etc., and their applications in calculations of QCD thermodynamics and phase structure, fluctuations and correlations of conserved charges, etc., see, e.g., \cite{Schaefer:2004en,Schaefer:2007pw,Skokov:2010wb,Herbst:2010rf, Skokov:2010uh, Karsch:2010hm,Morita:2011jva, Skokov:2011rq,Haas:2013qwp,Herbst:2013ail,Herbst:2013ufa,Fu:2015amv,Fu:2015naa,Fu:2016tey,Sun:2018ozp,Fu:2018qsk,Fu:2018swz,Wen:2018nkn,Wen:2019ruz,Yin:2019ebz,Hansen:2019lnf,Fu:2021oaw}.

\subsection{Flow equations}
\label{sec:flow}

Substituting the effective action in \Eq{eq:action} into the Wetterich equation in \Eq{eq:flow}, one readily obtains the flow equation of the effective potential as follows
\begin{align}
  \partial_t V_k(\rho)=&\frac{k^4}{4\pi^2} \bigg [\big(N^2_f-1\big) l^{(B,4)}_{0}(\bar{m}^{2}_{\pi,k},\eta_{\phi,k};T)\nonumber\\[2ex]
&+l^{(B,4)}_{0}(\bar{m}^{2}_{\sigma,k},\eta_{\phi,k};T)\nonumber\\[2ex]
&-4N_c N_f l^{(F,4)}_{0}(\bar{m}^{2}_{q,k},\eta_{q,k};T,\mu)\bigg]\,, \label{eq:flowV}
\end{align}
with the threshold functions $l^{(B,4)}_{0}$ and $l^{(F,4)}_{0}$ given in \Eq{eq:l0B} and \Eq{eq:l0F}, respectively. Here, the scale-dependent meson and quark masses read
\begin{align}
  \bar{m}^{2}_{\pi,k}&=\frac{V'_k(\rho)}{k^2Z_{\phi,k}}\,, \qquad \bar{m}^{2}_{\sigma,k}=\frac{V'_k(\rho)+2\rho V''_k(\rho)}{k^2 Z_{\phi,k}}\,,\label{eq:massmeson}\\[2ex]
  \bar{m}^{2}_{q,k}&=\frac{h_{y,k}^{2}\rho}{2k^2Z^{2}_{q,k}}\,,
\end{align}
which are RG invariant and dimensionless. 

The meson and quark anomalous dimensions in the threshold functions in \Eq{eq:flowV} are defined as follows
\begin{align}
  \eta_{\phi,k}&=-\frac{\partial_t Z_{\phi,k}}{Z_{\phi,k}}\,,\quad \eta_{q,k}=-\frac{\partial_t Z_{q,k}}{Z_{q,k}}\,, \label{}
\end{align}
where the meson anomalous dimension is obtained by projecting the flow equation in \Eq{eq:flow} onto the inverse pion propagator, to wit,
\begin{align}
  \eta_{\phi,k}(\rho)&=-\frac{1}{3Z_{\phi,k}}\delta_{ij}\frac{\partial}{\partial (|\bm{p}|^2)}\frac{\delta^2 \partial_t \Gamma_k}{\delta \pi_i(-p) \delta \pi_j(p)}\Bigg|_{\substack{p_0=0\\ \bm{p}=0}}\,,\label{eq:etaphi}
\end{align}
the explicit expression of which is presented in \Eq{eq:etaphi2}. Note that $\eta_{\phi,k}$ is dependent on the meson field via $\rho$. 

In comparison to the effects of the meson wave function renormalization on the chiral phase transition at finite temperature and density, it has been found that those of quark wave function renormalization and the running Yukawa coupling are relatively milder, see, e.g., \cite{Pawlowski:2014zaa,Fu:2015naa,Yin:2019ebz}. Therefore, in this work we adopt the simplification as follows
\begin{align}
  \eta_{q,k}&=0\,,\quad\quad \partial_t \bar h_{y,k}=0\,, \label{eq:Yukawa}
\end{align}
with the renormalized Yukawa coupling given in \Eq{eq:barhk}, and use two different truncations: one is the usual local potential approximation (LPA), where the mesonic anomalous dimension is vanishing as well, and the $k$-dependent term in \Eq{eq:action} is just the effective potential; the other is the truncation with the field-dependent mesonic anomalous dimension in \Eq{eq:etaphi} taken into account besides the potential, which is denoted as LPA$'$ in this work. Note that the notation LPA$'$ in literatures, e.g., \cite{Helmboldt:2014iya,Fu:2015naa}, usually stands for the truncation with a field-independent mesonic anomalous dimension which is, strictly speaking, different from the case in this work. 

%
\begin{figure}[t]
\includegraphics[width=0.5\textwidth]{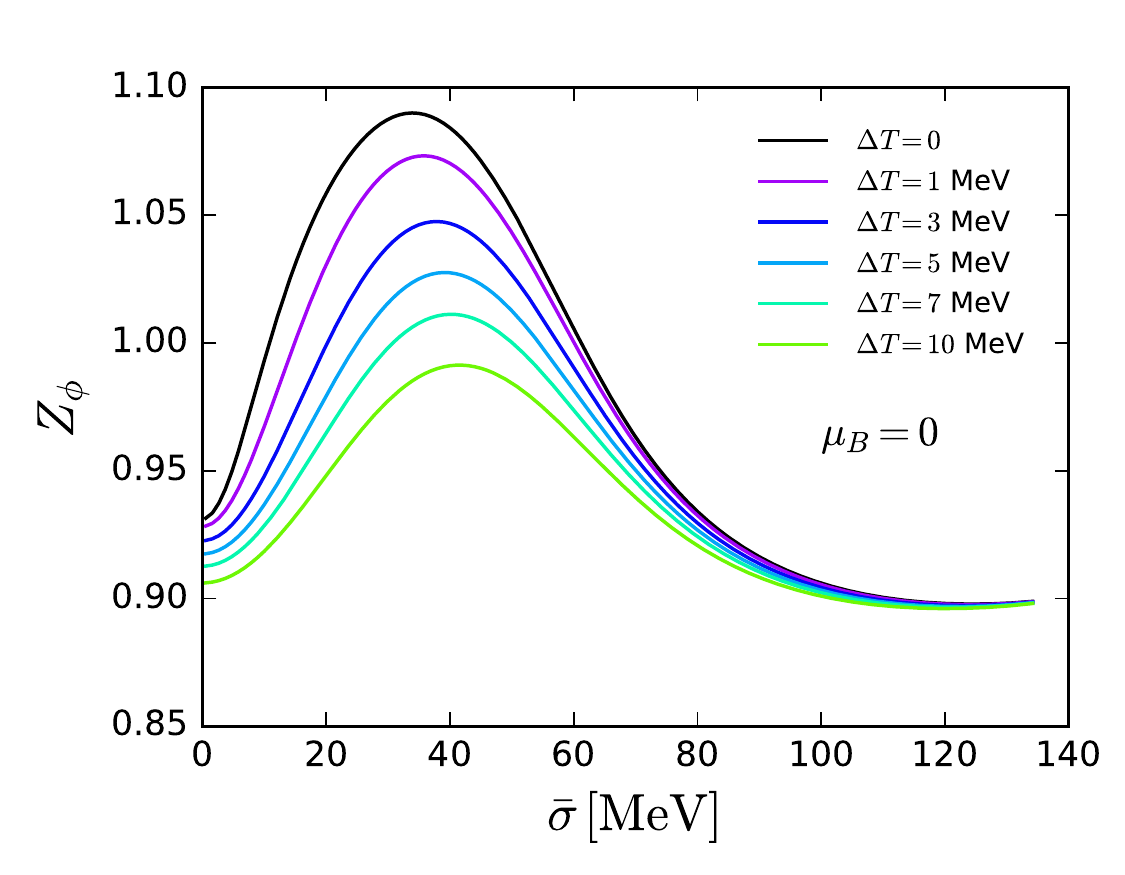}
\caption{Dependence of the mesonic wave function renormalization $Z_{\phi}$ on the order-parameter field $\bar \sigma$ at vanishing baryon chemical potential $\mu_B=0$ and several values of temperature $T=\Delta T+T_c$. See text for more details.}\label{fig:Zphi}
\end{figure}
%

As an illustrative example, we show the mesonic wave function renormalization $Z_{\phi}\equiv Z_{\phi,k=k_{\mathrm{IR}}}$  as a function of the renormalized sigma field $\bar \sigma=Z_{\phi}^{1/2} \sigma$ obtained in LPA$'$ in \Fig{fig:Zphi}, where $k_{\mathrm{IR}}$ is the RG scale in the IR limit, and one would has $k_{\mathrm{IR}} \rightarrow 0$ in principle, which, however, is impossible to realize in numerical calculations. In our calculation the value of $k_{\mathrm{IR}}$ is reduced as small as possible, and we find the convergence is obtained when $k_{\mathrm{IR}}=1$ MeV. Note that the mesonic wave function renormalization at the scale of UV cutoff $\Lambda$, see \sec{sec:phasediagram} in the following, is assumed to be identical to unity, i.e.,  $Z_{\phi,k=\Lambda}=1$. In \Fig{fig:Zphi}, we choose several values of temperature $T=\Delta T+T_c$ at and above the critical temperature that is $T_c=143.6$ MeV in the chiral limit and at vanishing $\mu_B$. One observes that with the increase of the temperature, the peak structure of $Z_{\phi}$ as a function of the renormalized sigma field $\bar \sigma$ becomes smoother.

\subsection{Chebyshev expansion of the effective potential}
\label{sec:ChebExpan}

%
\begin{figure*}[t]
\includegraphics[width=0.48\textwidth]{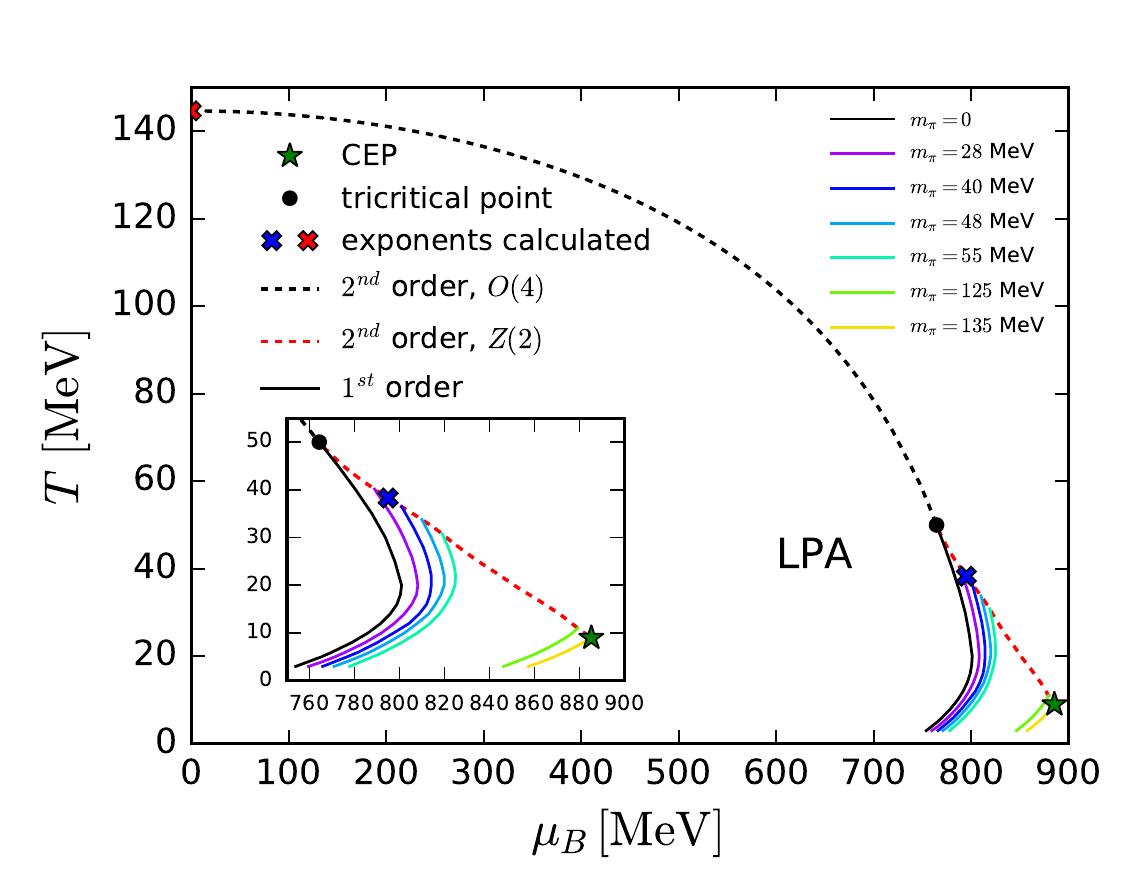}
\includegraphics[width=0.48\textwidth]{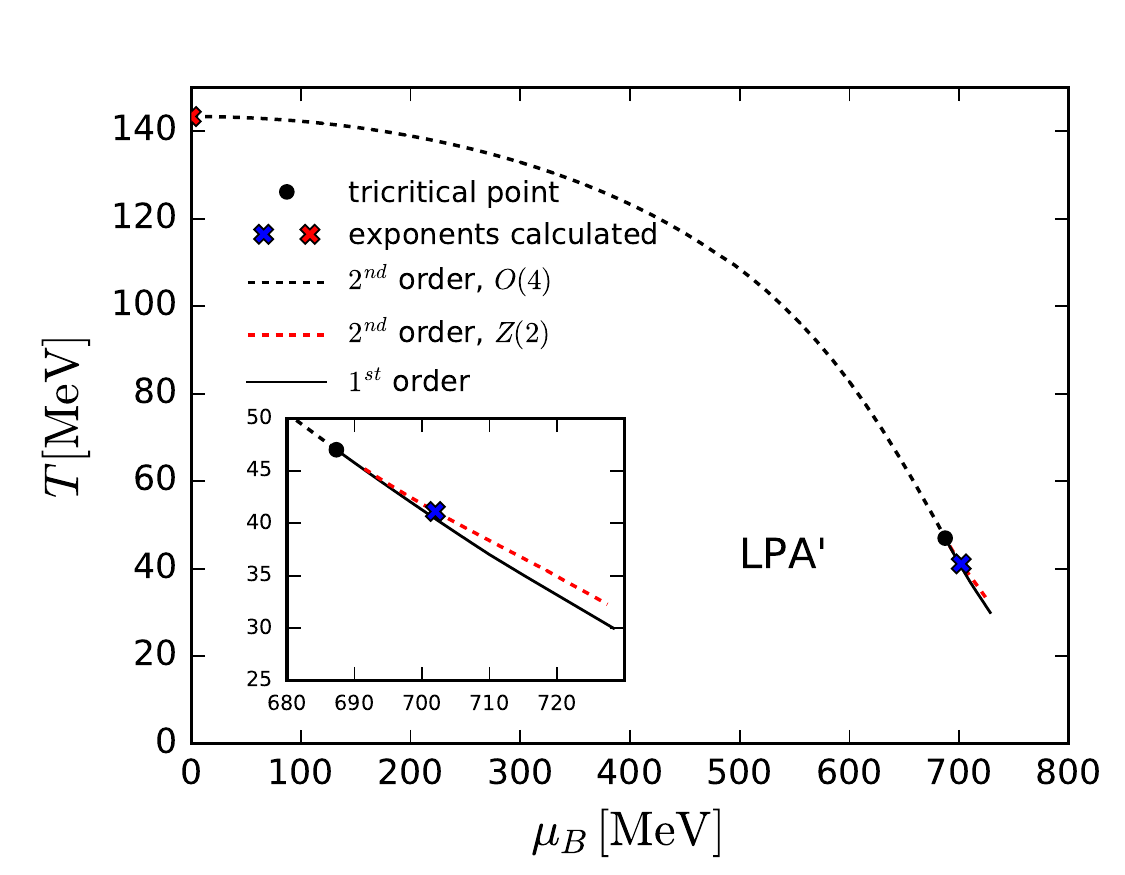}
\caption{Phase diagrams in the plane of $T$ and $\mu_B$, obtained in the quark-meson low energy effective theory within the fRG approach. Two truncations for the fRG calculations have been employed: one is the local potential approximation (LPA) and the other is that beyond the LPA, in which a field-dependent mesonic wave function renormalization is taken into account, i.e., the truncation LPA$'$, and see text for more details. The relevant results are presented in the left and right panels, respectively.\\
The black dashed lines in both panels denote the $O(4)$ chiral phase transition in the chiral limit, and the black circles indicate the location of the tricritical point. The solid lines of different colors in the left panel denote the first-order phase transitions with different pion masses in the vacuum, i.e. different values of $c$ in \Eq{eq:action}, and the solid one in the right panel is the first-order phase transition line in the chiral limit. The red dashed lines in both panels stand for line composed of critical end points (CEP) corresponding to continuously varying pion masses, which belong to the $Z(2)$ symmetry class. The star in the left panel indicates the location of CEP with physical pion mass. In both phase diagrams we use red and blue crosses to label the locations where critical exponents in \sec{sec:exponent} are calculated for the $O(4)$ and $Z(2)$ universality classes, respectively.}\label{fig:phasedia}
\end{figure*}
%

In this work we solve the flow equation in \Eq{eq:flowV} by expanding the effective potential as a sum of Chebyshev polynomials up to an order $N_v$, to wit,
\begin{align}
  \bar V_k(\bar \rho)&=\sum^{N_v}_{n=1} c_{n,k}T_n(\bar \rho)+\frac{1}{2}c_{0,k}\,,\label{eq:barVk}
\end{align}
with $\bar V_k(\bar \rho)=V_k(\rho)$, $\bar \rho=Z_{\phi,k} \rho$, where quantities with a bar denote renormalized variables. The Chebyshev polynomial $T_n(\bar \rho)$ is given in \Eq{eq:geneCheby}, and the superscript $[0,\bar \rho_{\mathrm{max}}]$ in \Eq{eq:geneCheby} denoting the interval of $\bar \rho$ is neglected for brevity here. Differentiating \Eq{eq:barVk} with respect to the RG time $t$ with $\rho$ fixed, one is led to
\begin{align}
  \partial_t\big|_{\rho} \bar V_k(\bar \rho)=&\sum^{N_v}_{n=1} \Big(\partial_t c_{n,k}-d_{n,k}\eta_{\phi,k}(\bar \rho)\bar \rho \Big)T_n(\bar \rho)\nonumber\\[2ex]
&+\frac{1}{2}\Big(\partial_t c_{0,k}-d_{0,k}\eta_{\phi,k}(\bar \rho)\bar \rho \Big)\,,\label{}
\end{align}
where we have used the Chebyshev expansion for the derivative of the effective potential as shown in \Eq{eq:chebyDeri} and coefficients $d_{n,k}$'s are the respective expanding coefficients. Employing the discrete orthogonality relation in \Eq{eq:ortho}  by summing up the $N+1$ zeros of $T_{N+1}(\bar \rho)$ in \Eq{eq:zeros2}, one arrives at
\begin{align}
  \partial_t c_{m,k}=&\frac{2}{N+1}\sum^{N}_{i=0}\Big(\partial_t\big|_{\rho} \bar V_k(\bar \rho_i)\Big)T_m(\bar \rho_i) \nonumber\\[2ex]
&+\frac{2}{N+1}\sum^{N_v}_{n=1}\sum^{N}_{i=0}d_{n,k}T_m(\bar \rho_i) T_n(\bar \rho_i)\eta_{\phi,k}(\bar \rho_i)\bar \rho_i \nonumber\\[2ex]
&+\frac{1}{N+1}d_{0,k}\sum^{N}_{i=0}T_m(\bar \rho_i)\eta_{\phi,k}(\bar \rho_i)\bar \rho_i \,,\label{eq:flowCoeff}
\end{align}
which is the flow equation for the expansion coefficients in \Eq{eq:barVk}.

\section{Phase diagram}
\label{sec:phasediagram}

It is left to specify the parameters in the LEFT, prior to presenting our calculated results. The UV cutoff of flow equations in the LEFT is chosen to be $\Lambda=500$ MeV, and the effective potential in \Eq{eq:action} at $k=\Lambda$ reads
\begin{align}
  V_\Lambda(\rho)&=\frac{\lambda_\Lambda}{2}\rho^2+\nu_\Lambda\rho\,,\label{}
\end{align}
with $\lambda_\Lambda=20$ and $\nu_\Lambda=0$ . The Yukawa coupling is $k$-independent as shown in \Eq{eq:Yukawa} and is given by $\bar h_{y}=6.4$. Concerning the Chebyshev expansion, we choose $N=81$ for the number of zeros and $N_v=21$ for the maximal order of Chebyshev polynomials. We have also checked that there is no difference when the value of $N_v$ is increased. Moreover, the upper bound of $\bar \rho$ is chosen to be $\bar \rho_{\mathrm{max}}=9\times 10^3\,\mathrm{MeV}^2$, well above the value of minimum of the potential in the IR. In the LPA, these values of parameters lead to the pion decay constant $f_\pi=$87 MeV and the constituent quark mass $m_q=$278.4 MeV in the vacuum and in the chiral limit. While if the explicit breaking strength of the chiral symmetry in \Eq{eq:action} is increased up to $c=1.85\times 10^{-3}\,(\mathrm{GeV})^3$, one obtains the physical pion mass $m_\pi=$138 MeV, as well as $f_\pi=$93 MeV and $m_q=$297.6 MeV in the vacuum. Note that in order to facilitate the comparison between the calculation with the truncation LPA and that with LPA$'$, we use the same values of parameters above in the LPA$'$ computation as in LPA.

In \Fig{fig:phasedia} we show the phase diagrams of LEFT in the $T\!-\!\mu_B$ plane, calculated within the fRG approach with the truncations LPA and LPA$'$, in the left and right panels, respectively. The black dashed lines in both panels denote the second-order $O(4)$ chiral phase transition of $N_f=2$ flavor in the chiral limit. The black circles indicate the location of the tricritical point, beyond which the second-order phase transition evolves into a discontinuous first-order one, which are shown by the solid lines. Note that the solid lines of different colors in the left panel denote the first-order phase transitions with different pion masses in the vacuum, i.e. different values of $c$ in \Eq{eq:action}, and in the right panel, we only give the first-order phase transition line in the chiral limit, since numerical calculations become quite difficult in the region of high $\mu_B$ and low $T$ with the truncation LPA$'$. The red dashed lines in both panels are the trajectories of the critical end points with the change of the strength of explicit chiral symmetry breaking $c$, which belong to the 3-$d$ $Z(2)$ Ising university class.

The critical temperature at vanishing baryon chemical potential is found to be $T_c=144$ MeV in LPA and 143
 MeV in LPA$'$ in the chiral limit. The tricritical point is located at $(T_{\mathrm{tri}},{\mu_B}_{\mathrm{tri}})_{_{\tiny{\mathrm{LPA}}}}=(50, 764)$ MeV in the LPA and $(T_{\mathrm{tri}},{\mu_B}_{\mathrm{tri}})_{_{\tiny{\mathrm{\mathrm{LPA}'}}}}=(47, 687)$ MeV in the LPA$'$, which are shown in the phase diagrams by the black circles. The location of CEP corresponding to the physical pion mass in the LPA, shown in the left panel of \Fig{fig:phasedia} by the star, is $(T_{_{\tiny{\mathrm{CEP}}}},{\mu_B}_{_{\tiny{\mathrm{CEP}}}})_{_{\tiny{\mathrm{LPA}}}}
  =(8, 885)$ MeV. In both phase diagrams in \Fig{fig:phasedia} we also use red and blue crosses to label the locations where critical exponents in \sec{sec:exponent} would be calculated for the 3-$d$ $O(4)$ and $Z(2)$ universality classes, respectively. The calculated points for the $O(4)$ and $Z(2)$ phase transition in the LPA are given by $(T_{_{O(4)}},{\mu_B}_{_{O(4)}})_{_{\tiny{\mathrm{LPA}}}}=(144, 0)$ MeV and $(T_{_{Z(2)}},{\mu_B}_{_{Z(2)}})_{_{\tiny{\mathrm{LPA}}}}=(38, 795)$ MeV, respectively; and the relevant values in the LPA$'$ read $(T_{_{O(4)}},{\mu_B}_{_{O(4)}})_{_{\tiny{\mathrm{LPA}'}}}=(143, 0)$ MeV and $(T_{_{Z(2)}},{\mu_B}_{_{Z(2)}})_{_{\tiny{\mathrm{LPA}'}}}=(41, 702)$ MeV.

\section{Critical behavior and critical exponents}
\label{sec:exponent}

A variety of scaling analysis has been performed for the $O(4)$ universality class, e.g., in the $O(N)$ model \cite{Toussaint:1996qr,Engels:1999wf,ParisenToldin:2003hq,Engels:2003nq,Braun:2007td,Engels:2009tv} and two-flavor quark-meson model \cite{Berges:1997eu,Schaefer:1999em,Bohr:2000gp,Stokic:2009uv}. The dynamics of a system in the critical regime near a second-order critical point is governed by long-wavelength fluctuations, and the correlation length tends to be divergent as the system moves towards the critical point. Critical exponents play a pivotal role in studies of the critical dynamics, which are independent of micro interactions, but rather universal for the same symmetry class, dimension of the system, etc., and see \cite{Stokic:2009uv,Braun:2007td} for more details. In the following, we follow the standard procedure and give our notations for the relevant various critical exponents.

To begin with, from the effective action in \Eq{eq:action} one readily obtains the thermodynamic potential density, which reads
\begin{align}
  \Omega\big(T,\mu_B,\,c\big)&=V_{k=0}(\rho)-c\sigma\,,\label{eq:omega}
\end{align}
where the order parameter field $\sigma\equiv\langle \sigma\rangle$ or $\rho=\sigma^2/2$ is on its equation of motion. We then introduce the reduced temperature and reduced external ``magnetic'' field as follows
\begin{align}
  t&=\frac{T-T_c}{T_0}\,,\qquad h=\frac{c}{c_0}\,, \label{eq:th}
\end{align}
where $T_c$ is the critical temperature, and they are normalized by $T_0$ and $c_0$, i.e., some appropriate values of $T$ and $c$. In the language of magnetization under an external magnetic field, the order parameter $\sigma$ here is just the corresponding magnetization density, i.e., $M\equiv \sigma$, and the explicit chiral symmetry breaking parameter $c$ is equivalent to the magnetic field strength $H\equiv c$. We will not distinguish them in the following any more. In the critical regime the thermodynamic potential in \Eq{eq:omega} is dominated by its singular part $f_s$, i.e.,
\begin{align}
  \Omega\big(t,h\big)&=f_s(t,h)+f_{reg}(t,h)\,,\label{eq:scalingfunc}
\end{align}
where the second term on the r.h.s. is the regular one, and the notation for the baryon chemical potential is suppressed. In what follows we adopt the notations in \cite{Braun:2010vd}, and the scaling function $f_s(t,h)$ on the r.h.s. of \Eq{eq:scalingfunc} satisfies the scale relation to leading order, viz.
\begin{align}
  f_s(t,h)&=\ell^{-d}f_s(t\,\ell^{y_t},\,h\,\ell^{y_h})\,,\label{eq:scalingrelation}
\end{align}
where $\ell$ is a dimensionless rescaling factor. The scaling function in \Eq{eq:scalingrelation} leads us to a variety of relations for various critical exponents \cite{Berges:1997eu,Tetradis:2003qa,Schaefer:1999em,Braun:2007td}, e.g.,
\begin{align}
  y_t&=\frac{1}{\nu}\,,\quad\!\! y_h=\frac {\beta\delta}{\nu}\,,\quad\!\! \beta=\frac{\nu}{2} (d-2+\eta)\,,\quad\!\! \gamma=\beta(\delta-1)\,, \nonumber\\[2ex]
\gamma&=(2-\eta)\nu\,,\quad\delta=\frac {d+2-\eta}{d-2+\eta}\,,\quad \nu d=\beta(1+\delta)\,,\label{eq:relationExpo}
\end{align}
with the spacial dimension $d$. The critical exponents $\beta$ and $\delta$ describe the critical behavior of the order parameter in the direction of $t$ or $h$, respectively, i.e.,
\begin{align}
  M(t,h=0)&\sim (-t)^{\beta}\quad \mathrm{with}\quad t<0\,,\label{eq:beta}\\[2ex]
  M(t=0,h)&\sim h^{1/\delta}\,.\label{eq:delta}
\end{align}
The exponent $\gamma$ is related to the susceptibility of order parameter $\chi$, and $\nu$ to the correlation length $\xi$, which reads
\begin{align}
  \chi&\sim |t|^{-\gamma}\,,\quad \mathrm{and}\quad \xi\sim |t|^{-\nu}\,,\label{eq:defiGamNu}
\end{align}

The scaling relation in \Eq{eq:scalingrelation} allows us to readily obtain the critical behavior for various observables. For instance, the order parameter and its susceptibilities read
\begin{align}
  M&=-\frac {\partial f_s}{\partial H}\,,\quad \chi_{\sigma}=\frac {\partial M}{\partial H}\,,\quad\chi_{\pi}=\frac {M}{H}\,,\label{eq:Mchi}
\end{align}
where $\chi_{\sigma}\equiv \chi_l$ and $\chi_{\pi}\equiv \chi_t$ are also called as the longitudinal and transverse susceptibilities, respectively.
 
Choosing an appropriate value of the rescaling factor such that $h\,\ell^{y_h}=1$ in \Eq{eq:scalingrelation}, one is led to
\begin{align}
  f_s(t,h)&=h^{d/y_h} f_s(z,1)\,,\label{eq:scalingrelation2}
\end{align}
with the scaling variable $z=t/h^{1/(\beta\delta)}$. Inserting \Eq{eq:scalingrelation2} into the first equation in \Eq{eq:Mchi}, one arrives at
\begin{align}
  M&=h^{1/\delta} f(z)\,,\label{eq:Mfz}
\end{align}
where we have introduced 
\begin{align}
  f(z)&\equiv \frac{1}{H_0}\Big[\frac{z}{\beta\delta}\frac{\partial f_s(z,1)}{\partial z}-\frac{d\nu}{\beta\delta}f_s(z,1)\Big]\,,\label{eq:fz}
\end{align}
which is a scaling function dependent only on $z$. With appropriate values of $H_0$ and $T_0$ in \Eq{eq:th}, it can be shown that the scaling function in \Eq{eq:fz} has the properties $f(0)=1$ and $f(z)\simeq(-z)^\beta$ with $z\rightarrow -\infty$ \cite{Braun:2010vd}.

Consequently, it is straightforward to express the longitudinal and transverse susceptibilities in \Eq{eq:Mchi} in terms of the scaling function $f(z)$, to wit,
\begin{align}
  \chi_{\sigma}&=\frac{1}{H_0}h^{1/\delta-1}f_\chi(z)\,,\label{}
\end{align}
with
\begin{align}
  f_\chi(z)&\equiv \frac{1}{\delta}\Big[f(z)-\frac z\beta f'(z)\Big]\,,\label{}
\end{align}
and 
\begin{align}
  \chi_{\pi}&=\frac{1}{H_0}h^{1/\delta-1}f(z)\,.\label{}
\end{align}

Alternative to the choice of $h\,\ell^{y_h}=1$ in \Eq{eq:scalingrelation}, one can also employ $t\,\ell^{y_t}=1$, which is equivalent to the Widom-Griffiths parametrization \cite{Widom:1965,Griffiths:1967zza} of the equation of state by means of the scaling variables, as follows
\begin{align}
  x&\equiv\frac{t}{M^{1/\beta}}\,,\qquad y\equiv\frac{h}{M^\delta}\,,\label{eq:widomgriff}
\end{align}
which are obviously related to the other parametrization by the relations which read
\begin{align}
  z&=\frac{x}{y^{1/(\beta\delta)}}\,,\qquad f(z)=\frac{1}{y^{1/\delta}}\,.\label{eq:xyzf}
\end{align}
Hence the scaling function $y(x)$ has the properties $y(0)=1$ and $y(-1)=0$. In the same way, one readily obtains the expressions of susceptibilities in this parametrization, which read
\begin{align}
  \chi_{\sigma}&=\frac{1}{H_0 M^{\delta-1}}\Big[\delta y(x)-\frac 1\beta xy'(x)\Big]^{-1}\,,\label{eq:chisigWG}\\[2ex]
  \chi_{\pi}&=\frac{1}{H_0 M^{\delta-1}}\frac{1}{y}\,.\label{eq:chipiWG}
\end{align}
%


\subsection{Order parameter}
\label{sec:scaorder}

%
\begin{figure*}[t]
\includegraphics[width=1.0\textwidth]{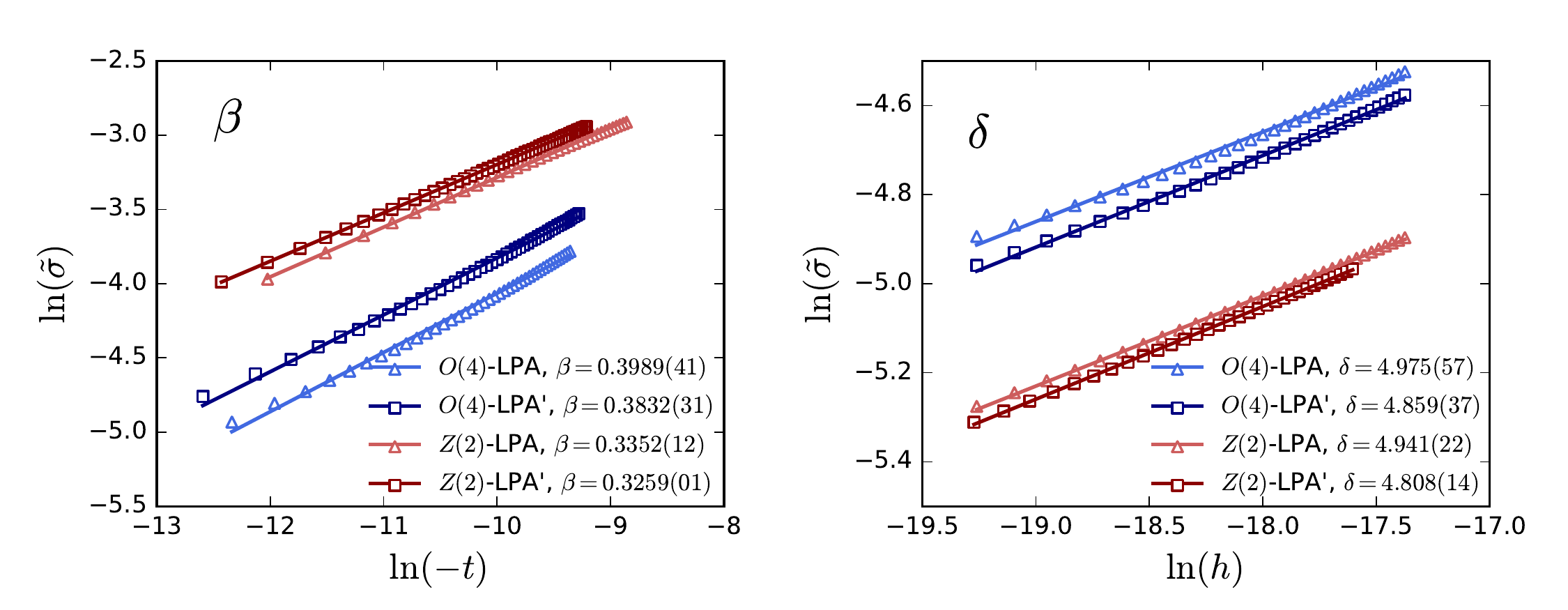}
\caption{Logarithm of the reduced order parameter $\tilde{\sigma}$ in \Eq{eq:reduSig} as a function of $\ln(-t)$ (left panel) or $\ln(h)$ (right panel) for the second-order $O(4)$ and $Z(2)$ phase transitions with truncations LPA and LPA$'$, where the phase transition points are chosen to be the locations of the red and blue crosses in the phase diagrams in \Fig{fig:phasedia} for the $O(4)$ and $Z(2)$ universality classes, respectively. The solid lines represent linear fits to the calculated discrete data points, from which values of the critical exponents $\beta$ and $\delta$ are extracted.}\label{fig:betadelta}
\end{figure*}
%

%
\begin{table*}[t]
  \begin{center}
  \begin{tabular}{cccccc}
    \hline\hline & & & & & \\[-2ex]   
    $T_c-T$ (MeV) & \big($10^{-4}$, $5\!\times\!10^{-3}$\big) & \big($10^{-2}$, 0.1\big) & \big(0.1, 0.5 \big) & \big(0.5, 1\big) & \big(1, 5\big)\\[1ex]
    \hline & & & & & \\[-2ex]
    $\beta^{^{O(4)}}_{_{\mathrm{LPA}}}$ & 0.3989(41) & 0.5164(65) & 0.4374(36) & 0.4077(44) &0.3921(43) \\[1ex]
    $\beta^{^{Z(2)}}_{_{\mathrm{LPA}}}$ & 0.3352(12) & 0.2830(26) & 0.2724(18) & 0.2689(17) &0.247(17) \\[1ex]    
    \hline\hline
  \end{tabular}
  \caption{Values of the critical exponent $\beta$ extracted from different ranges of temperature, which are denoted by their distances to the corresponding critical temperature, i.e., $T_c-T$. The calculations are performed with the truncation LPA, and the phase transition points are chosen to be the locations of the red and blue crosses in the phase diagrams in \Fig{fig:phasedia} for the $O(4)$ and $Z(2)$ universality classes, respectively.} 
  \label{tab:beta-critreg}
  \end{center}\vspace{-0.5cm}
\end{table*}
%

The flow equation of effective potential in \Eq{eq:flowV} is solved by the use of the Chebyshev expansion as discussed in \sec{sec:ChebExpan}, i.e., evolving the flow equations of the expansion coefficients in \Eq{eq:flowCoeff} from the UV cutoff $\Lambda$ to the infrared limit $k\rightarrow 0$, and then the expectation value of the order parameter $\sigma$ is determined by minimizing the thermodynamic potential in \Eq{eq:omega}. Note that two different truncations, i.e., LPA and LPA$'$ as shown in \sec{sec:flow}, are employed in the calculations.

The critical exponents $\beta$ and $\delta$ are given in Eqs. (\ref{eq:beta}) and (\ref{eq:delta}), which are related to the scaling behavior of the order parameter as the phase transition is approached towards in the temperature or external field direction, respectively. Note, however, that in the case of $Z(2)$ phase transition as indicated by the blue cross in the phase diagram in \Fig{fig:phasedia}, the order parameter should be modified slightly and we introduce the reduced order parameter which reads
\begin{align}
  \tilde{\sigma}&=\frac{\sigma-\sigma'}{f_\pi}\,,\label{eq:reduSig}
\end{align}
where $f_\pi$ is the pion decay constant in the vacuum and $\sigma'$ is the expectation value of sigma field at the phase transition point, which is nonvanishing on the red dashed lines of $Z(2)$ in the phase diagrams in \Fig{fig:phasedia}. Correspondingly, the reduced external field in \Eq{eq:th} is modified into
\begin{align}
  h&=\frac{c-c'}{c_0}\,,\label{eq:reduh}
\end{align}
where $c'$ is the $\sigma'$-related external field on the $Z(2)$ phase transition line. Notice that both $c'$ and $\sigma'$ are vanishing on the $O(4)$ phase transition line, viz., the black dashed lines in \Fig{fig:phasedia}. In our calculations below, the normalized external field strength $c_0$ in \Eq{eq:reduh} is chosen to be the value corresponding to the physical pion mass, and the normalized temperature in \Eq{eq:th} is to be the critical one $T_0=T_c$.

In \Fig{fig:betadelta} we show the log-log plots of the reduced order parameter $\tilde{\sigma}$ versus the reduced temperature $-t$ or external field $h$ for the second-order $O(4)$ and $Z(2)$ phase transitions. The calculations are performed in the quark-meson LEFT with the fRG in both LPA and LPA$'$. The phase transition points are chosen to be the locations of the red and blue crosses in the phase diagrams in \Fig{fig:phasedia} for the $O(4)$ and $Z(2)$ universality classes, respectively. A linear relation is used to fit the calculated discrete data points in \Fig{fig:betadelta}, and as shown in \Eq{eq:beta} and \Eq{eq:delta}, one could extract the values of the critical exponents $\beta$ and $\delta$ from the slope of these linear curves. This leads us to 
\begin{align}
  \beta^{^{O(4)}}_{_{\mathrm{LPA}}}&=0.3989(41)\,,\qquad \beta^{^{O(4)}}_{_{\mathrm{LPA}'}}=0.3832(31)\,,\label{eq:betaO4}
\end{align}
for the $O(4)$ universality class in LPA and LPA$'$, respectively. In the case of the $Z(2)$ universality class, one arrives at
\begin{align}
  \beta^{^{Z(2)}}_{_{\mathrm{LPA}}}&=0.3352(12)\,,\qquad \beta^{^{Z(2)}}_{_{\mathrm{LPA}'}}=0.3259(01)\,.\label{}
\end{align}
In the same way, the values of $\delta$ are obtained as follows
\begin{align}
  \delta^{^{O(4)}}_{_{\mathrm{LPA}}}&=4.975(57)\,,\qquad \delta^{^{O(4)}}_{_{\mathrm{LPA}'}}=4.859(37)\,,\label{eq:deltaO4}\\[2ex]
  \delta^{^{Z(2)}}_{_{\mathrm{LPA}}}&=4.941(22)\,,\qquad \delta^{^{Z(2)}}_{_{\mathrm{LPA}'}}=4.808(14)\,.\label{}
\end{align}
It is found that the critical exponents $\beta$ and $\delta$ of the $O(4)$ and $Z(2)$ phase transitions in 3-$d$ systems calculated in this work are consistent with previous results, e.g., Monte Carlo simulation of spin model \cite{Kanaya:1994qe} and $d=3$ expansion for $Z(2)$\cite{ZinnJustin:1999bf}. Comparing the relevant results in LPA and LPA$'$, one observes that both $\beta$ and $\delta$ obtained in LPA$'$ are slightly smaller than those in LPA.





\subsection{Preliminary assessment of the size of the critical region}
\label{sec:critical-region}

It is well known that critical exponents are universal for the same universality classes. The size of the critical region is, however, non-universal and depends on the interactions and other details of system concerned. Furthermore, there has been a longstanding debate on the size of the critical region in QCD. Lattice QCD simulations show that the chiral condensate, i.e., the order parameter in \Eq{eq:Mfz}, for physical quark masses are well described by \Eq{eq:Mfz} plus a small analytic regular term \cite{Ejiri:2009ac,Kaczmarek:2011zz,Ding:2019prx}, which, in another word, implies that the size of the critical regime of QCD is large enough, such that QCD with physical quark mass is still in the chiral critical regime. On the contrary, it is found in \cite{Braun:2007td,Braun:2010vd,Klein:2017shl} that the pion mass required to observe the scaling behavior is very small, at least one order of magnitude smaller than the physical pion mass. Moreover, it is also found that the critical region around the CEP in the QCD phase diagram is very small \cite{Schaefer:2006ds}. In \tab{tab:beta-critreg} we present the values of the critical exponent $\beta$ extracted from different ranges of temperature. One observes that when the temperature range is away from the critical temperature larger than $0.01$ MeV, the value of $\beta$ deviates from its universal value pronouncedly. This applies for both the $O(4)$ and $Z(2)$ universality classes. Given the systematic errors in the computation of this work, one could safely conclude that our calculation indicates that the critical region in the QCD phase diagram is probably very small, and it is smaller than 1 MeV in the direction of temperature.

\subsection{Chiral susceptibility}
\label{sec:scasuscep}

%
\begin{figure}[t]
\includegraphics[width=0.5\textwidth]{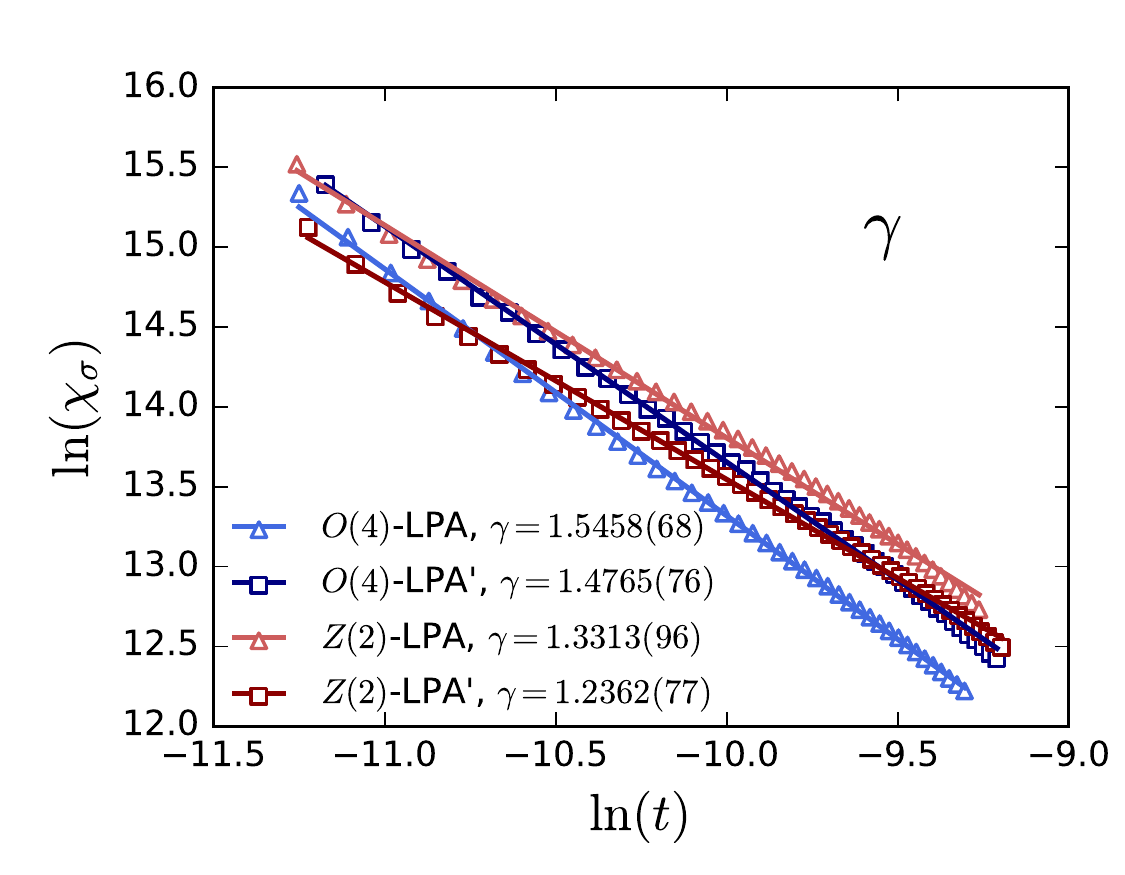}
\caption{Logarithm of the longitudinal susceptibility $\chi_{\sigma}$ as a function of $\ln(t)$ in the chiral symmetric phase. The calculation is done in the quark-meson LEFT within the fRG approach with truncations LPA and LPA$'$, where the phase transition points are chosen to be at the locations of the red and blue crosses in the phase diagrams in \Fig{fig:phasedia} for the $O(4)$ and $Z(2)$ universality classes, respectively. The solid lines represent linear fits to the calculated discrete data points, from which value of the critical exponent $\gamma$ is extracted.}\label{fig:gama}
\end{figure}
%

%
\begin{figure*}[t]
\includegraphics[width=1.0\textwidth]{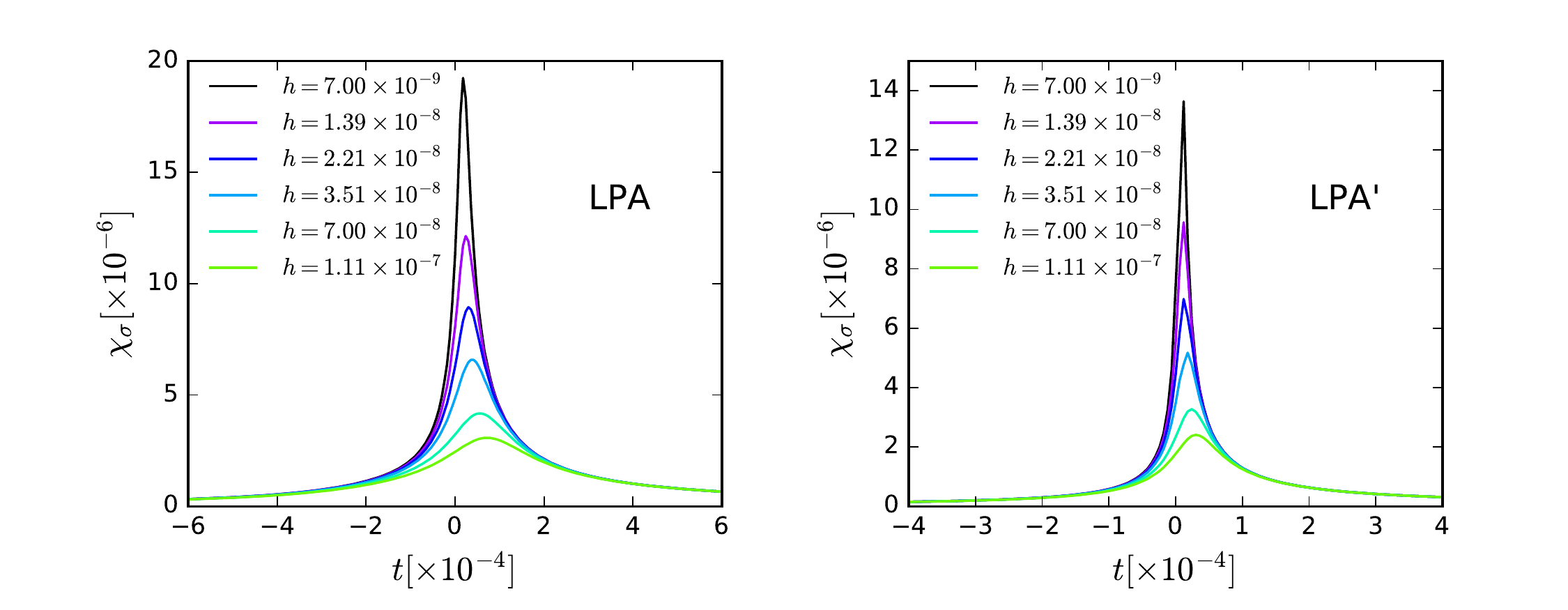}
\caption{Longitudinal susceptibility of the order parameter $\chi_\sigma$ as a function of the reduced temperature $t$ with several different values of the reduced external field $h$, calculated in the LPA (left panel) and LPA$'$ (right panel). The phase transition is chosen to be near the location of the red cross in the phase diagrams in \Fig{fig:phasedia} for the $O(4)$ symmetry universality class.}\label{fig:chisigma}
\end{figure*}
%

%
\begin{figure*}[t]
\includegraphics[width=1.0\textwidth]{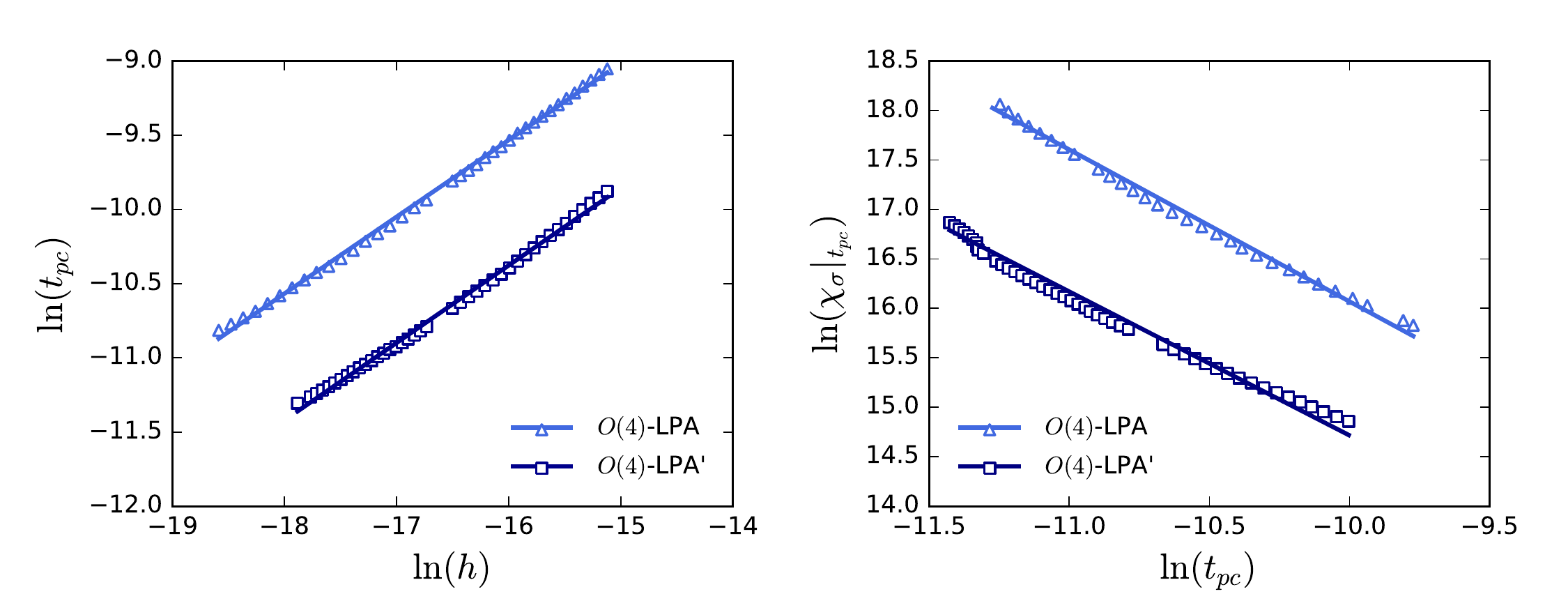}
\caption{Left panel: logarithm of the reduced pseudo-critical temperature $t_{pc}$, defined by the peak of the susceptibility $\chi_\sigma$ as shown in \Fig{fig:chisigma}, as a function of the logarithm of the reduced external field strength $h$. Right panel: logarithm of the peak height of the susceptibility, $\chi_\sigma\big|_{t_{pc}}$, versus the logarithm of the reduced pseudo-critical temperature.\\
Calculations are done within the fRG approach with the truncations LPA and LPA$'$. The phase transition is chosen to be near the location of the red cross in the phase diagrams in \Fig{fig:phasedia} for the $O(4)$ symmetry universality class.}\label{fig:Tpc}
\end{figure*}
%

%
\begin{figure*}[t]
\includegraphics[width=1.\textwidth]{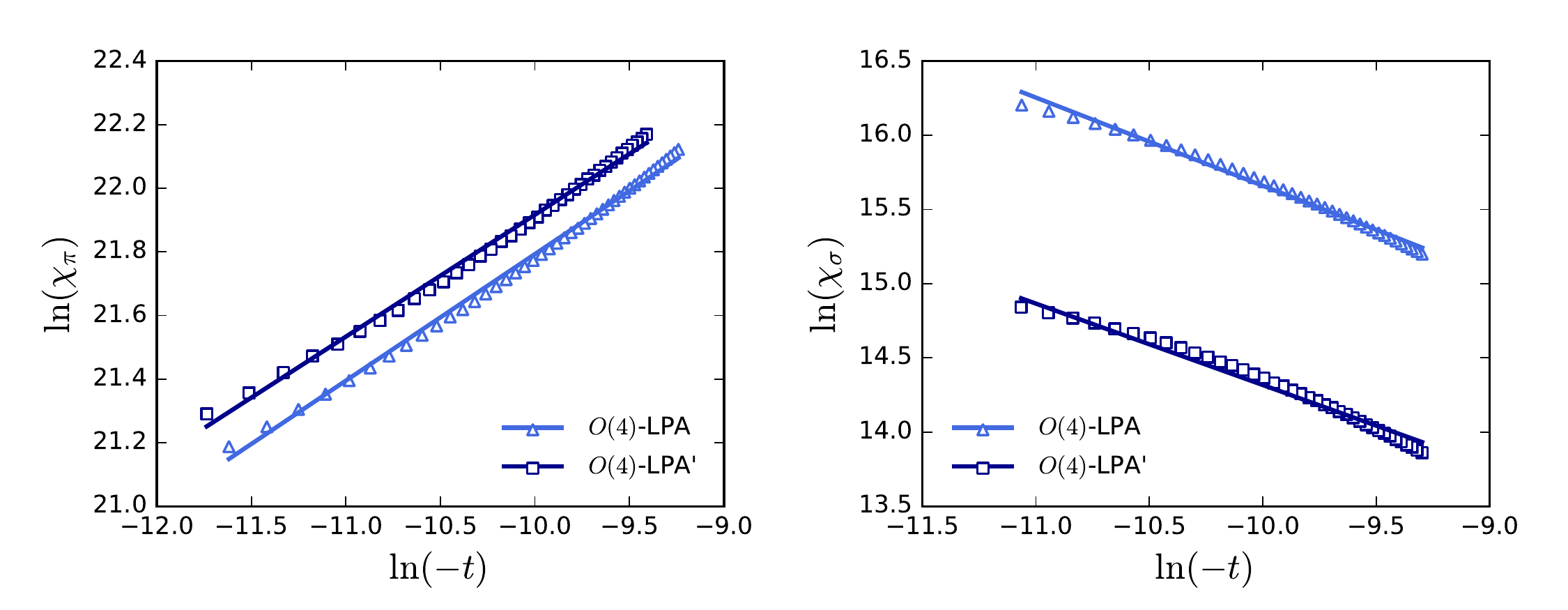}
\caption{Logarithms of the transverse (left panel) and longitudinal (right panel) susceptibilities as functions of the logarithm of $-t$ with a fixed value of the reduced external field $h=8.4\times10^{-9}$ in the chiral broken phase near the coexistence line. Calculations are performed within the fRG approach with the truncations LPA and LPA$'$. The phase transition is chosen to be near the location of the red cross in the phase diagrams in \Fig{fig:phasedia} for the $O(4)$ symmetry universality class, where the baryon chemical potential is vanishing.}\label{fig:chipisigcoex}
\end{figure*}
%

%
\begin{figure*}[t]
\includegraphics[width=1.0\textwidth]{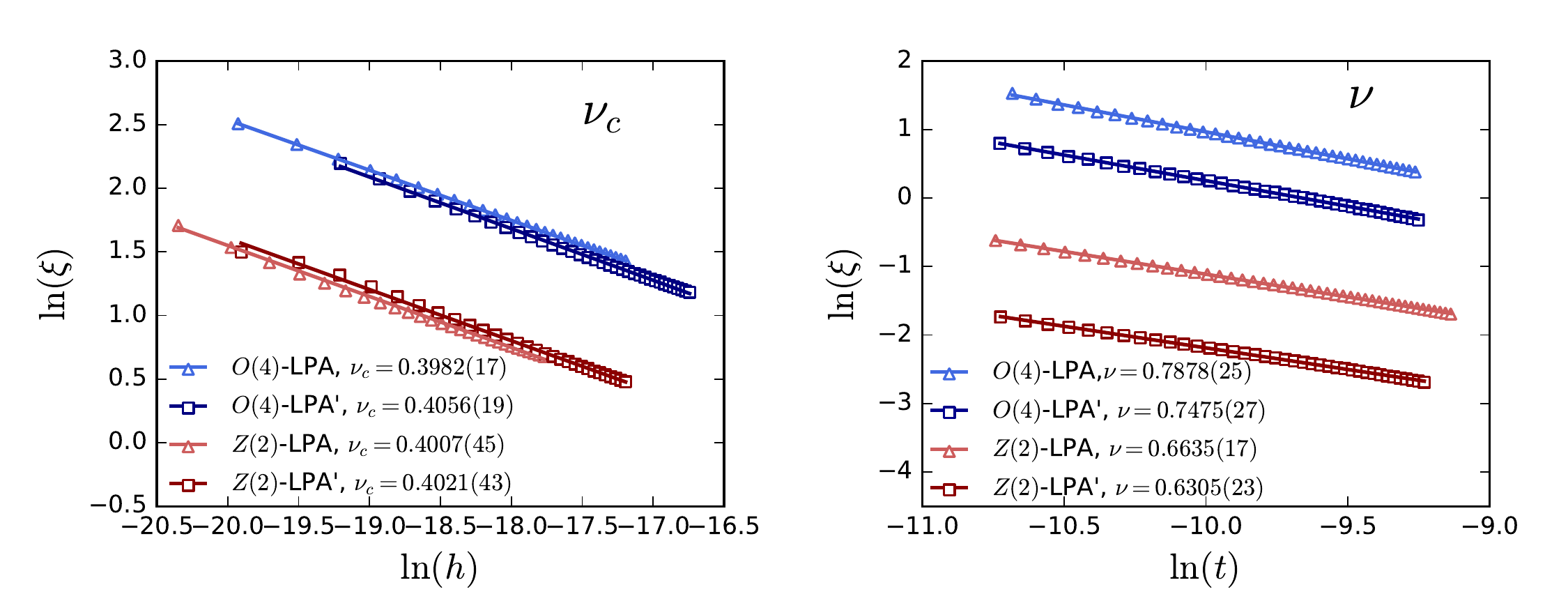}
\caption{Left panel: logarithm of the correlation length as a function of the logarithm of the reduced external field strength with $t=0$. Right panel: logarithm of the correlation length as a function of the logarithm of the reduced temperature with $h=0$.\\
Both calculations are performed in the quark-meson LEFT within the fRG approach with truncations LPA and LPA$'$, where the phase transition points are chosen to be at the locations of the red and blue crosses in the phase diagrams in \Fig{fig:phasedia} for the $O(4)$ and $Z(2)$ universality classes, respectively. The solid lines represent linear fits to the calculated discrete data points, from which values of the critical exponent $\nu_c$ and $\nu$ are yielded.}\label{fig:nucnu}
\end{figure*}
%

According to \Eq{eq:widomgriff}, the reduced order parameter reads
\begin{align}
  \tilde{\sigma}&\sim h^{1/\delta} y(x)^{-1/\delta}\,.\label{eq:sigy}
\end{align}
Moreover, it has been shown in \cite{Griffiths:1967zza}  that given $x>0$ and $M>M_0$ for some value $M_0$ in \Eq{eq:widomgriff}, the scaling function can be expanded as 
\begin{align}
  y(x)&=\sum_{n=1}^\infty c_n x^{\gamma-2\beta(n-1)}\nonumber\\[2ex]
        &=x^\gamma \big(c_1+c_2 x^{-2\beta}+c_3 x^{-4\beta}+\dots\big)\,.\label{eq:yExpan}
\end{align}
Inserting the leading term in \Eq{eq:yExpan} into \Eq{eq:sigy} and utilizing the relation $\gamma=\beta(\delta-1)$ as shown in \Eq{eq:relationExpo}, one is led to the reduced order parameter with $t>0$ and $h\rightarrow 0$, which reads
\begin{align}
  \tilde{\sigma}&\sim t^{-\gamma} h\,.\label{eq:sigth}
\end{align}
Consequently, the longitudinal and transverse susceptibilities of the order parameter as defined in \Eq{eq:Mchi} are readily obtained as follows
\begin{align}
  \chi_\sigma&=\chi_\pi\sim t^{-\gamma}\,,\label{}
\end{align}
which is in agreement with \Eq{eq:defiGamNu} in the limit $h\rightarrow 0$ and in the symmetric phase, as it should be. Equation~(\ref{eq:sigth}) also allows us to extract the value of the exponent $\gamma$, by directly investigating the scaling relation of $\tilde{\sigma}$ and $t$ in the chiral symmetric phase with a fixed, small value of $h$. In \Fig{fig:gama} we show the logarithm of the longitudinal susceptibility $\chi_{\sigma}$ versus that of the reduced temperature, where $h=3.5\times10^{-10}$ is chosen in the calculations. We have checked that this value of $h$ is small enough to make sure that the value of $\gamma$ obtained from the linear fit of $\ln (\chi_{\sigma})$-$\ln(t)$ is convergent. In the same way, the flow equations of fRG are resolved with two truncations LPA and LPA$'$, and the phase transition points are chosen to be at the locations of the red and blue crosses in the phase diagrams in \Fig{fig:phasedia} for the $O(4)$ and $Z(2)$ universality classes, respectively. The values of the exponent $\gamma$ are obtained as follows

\begin{align}
 \gamma^{^{O(4)}}_{_{\mathrm{LPA}}}&=1.5458(68)\,,\qquad \gamma^{^{O(4)}}_{_{\mathrm{LPA}'}}=1.4765(76)\,,\label{eq:gammaO4}\\[2ex]
  \gamma^{^{Z(2)}}_{_{\mathrm{LPA}}}&=1.3313(96)\,,\qquad \gamma^{^{Z(2)}}_{_{\mathrm{LPA}'}}=1.2362(77)\,.\label{}
\end{align}
Once more, one observes that these values, in particular those obtained in the LPA$'$, are in good agreement with the values of $\gamma$ for the $O(4)$ and $Z(2)$ symmetry universality classes, respectively; see, e.g., \cite{Kanaya:1994qe,ZinnJustin:1999bf}.
 
In \Fig{fig:chisigma} the longitudinal susceptibility of the order parameter $\chi_\sigma$, as shown in \Eq{eq:Mchi}, is depicted versus the reduced temperature with several different values of the reduced external field. Here we only focus on the case of $O(4)$ symmetry, and thus choose the phase transition to be near the location of the red cross in the phase diagrams in \Fig{fig:phasedia}, i.e., the phase transition with vanishing baryon chemical potential. When the external field $h$ that breaks the chiral symmetry explicitly is nonzero, the second-order phase transition becomes a continuous crossover, as shown in \Fig{fig:chisigma}. One can define a pseudo-critical temperature $T_{pc}$, which is the peak position of the curve $\chi_\sigma$ versus $T$, and thus the reduced pseudo-critical temperature reads
\begin{align}
  t_{pc}&=\frac{T_{pc}-T_c}{T_c}\,.\label{}
\end{align}
One observes from \Fig{fig:chisigma} that with the increasing $h$, the peak height of the susceptibility decreases and the pseudo-critical temperature $t_{pc}$ increases. The rescaling relation between $t_{pc}$ and $h$ as well as that between the peak height of $\chi_\sigma$ and $t_{pc}$ reads
\begin{align}
  t_{pc}&\sim h^{1/(\gamma+\beta)}\,,\qquad \chi_\sigma\big|_{t_{pc}}\sim t_{pc}^{-\gamma}\,,\label{eq:scalingtpch}
\end{align}
and see, e.g., \cite{Pelissetto:2000ek} for more details.

In \Fig{fig:Tpc} we show the logarithm of the reduced pseudo-critical temperature versus the logarithm of the reduced external field strength, and the logarithm of the peak height of the susceptibility versus the logarithm of the reduced pseudo-critical temperature in the left and right panels, respectively. The phase transition is also chosen to be near the location of the red cross in the phase diagrams in \Fig{fig:phasedia} for the $O(4)$ symmetry universality class, where the baryon chemical potential is vanishing. Linear fitting to the calculated discrete data in \Fig{fig:Tpc} yields $\beta=0.403(19)$ and $\gamma=1.543(15)$ for the LPA, and $\beta=0.405(22)$ and $\gamma=1.454(17)$ for the LPA$'$, which are in agreement with the relevant values in \Eq{eq:betaO4} and \Eq{eq:gammaO4} within errors for the $O(4)$ second-order phase transition in 3-$d$ space. In turn, the agreement of critical exponents obtained from different scaling relations also provides us with the necessary check for the inner consistency of computations. Note, however, that the critical exponents $\beta$ and $\gamma$ determined from the scaling relations in \Eq{eq:scalingtpch} are significantly less accurate than those in \Eq{eq:betaO4} and \Eq{eq:gammaO4}.

As another check for the consistency, we consider the susceptibilities in the chiral broken phase near the coexistence line, i.e., $x=-1$, with $t<0$ and $h\rightarrow 0$. Inserting \Eq{eq:sigy} into Eqs.~(\ref{eq:chisigWG})~(\ref{eq:chipiWG}), one is led to
\begin{align}
  \chi_{\sigma}&\sim h^{1/\delta-1}\frac{\beta y(x)^{1-1/\delta}}{\beta \delta y(x)- xy'(x)}\,,\label{eq:chisigcoex}\\[2ex]
  \chi_{\pi}&\sim h^{1/\delta-1}y(x)^{-1/\delta}\,.\label{}
\end{align}
when the system is near the coexistence line, one has $x\rightarrow-1$ and $y\sim h/(-t)^{\beta\delta}$. Hence, the transverse susceptibility is readily obtained as follows
\begin{align}
  \chi_{\pi}&\sim h^{-1} (-t)^\beta\,.\label{eq:chipicoex2}
\end{align}
In order to obtain a similar expression for the longitudinal susceptibility, one needs further information on the equation of state $y(x)$. As the system is located in the broken phase near the coexistence line, the dynamics is dominated by  Goldstone modes, which are massless in the chiral limit. The relevant critical behavior in this regime is governed by a Gaussian fixed point, and thus the corresponding exponents are as same as values of mean fields \cite{Wallace:1975vi,Brezin:1972se}, which leaves us with 
\begin{align}
 y(x)&\sim (1+x)^2\,,\qquad\mathrm{for}\qquad x\rightarrow-1\,,\label{eq:eoscoex}
\end{align}
and see, e.g., \cite{Braun:2007td,Stokic:2009uv} for more relevant discussions. Substituting equation above into \Eq{eq:chisigcoex}, one arrives at
\begin{align}
  \chi_{\sigma}&\sim h^{-1/2} (-t)^{\beta-(\beta\delta/2)}\,.\label{eq:chisigcoex2}
\end{align}
As Eqs.~(\ref{eq:chipicoex2})~(\ref{eq:chisigcoex2}) show, the transverse and longitudinal susceptibilities are proportional to the external field with different powers in the broken phase, i.e., $-1$ and $-1/2$ for the former and latter, respectively.

In \Fig{fig:chipisigcoex} we show $\ln(\chi_{\pi})$ and $\ln(\chi_{\sigma})$ versus $\ln(-t)$ with a fixed value of the reduced external field $h=8.4\times10^{-9}$ in the chiral broken phase near the coexistence line. Similarly, here we only consider the phase transition of $O(4)$ symmetry with $\mu_B=0$ in the phase diagrams in \Fig{fig:phasedia}. As shown in Eqs.~(\ref{eq:chipicoex2})~(\ref{eq:chisigcoex2}), the ratios of the linear fitting to $\ln(\chi_{\pi})$-$\ln(-t)$ and $\ln(\chi_{\sigma})$-$\ln(-t)$ are just the values of $\beta$ and $\beta-(\beta\delta/2)$, respectively. Consequently, one arrives at $\beta=0.3979(41)$ and $\delta=4.984(74)$ in LPA and $\beta=0.3832(54)$ and $\delta=4.86(10)$ in LPA$'$, which agree very well with the relevant values in \Eq{eq:betaO4} and \Eq{eq:deltaO4}.

\subsection{Correlation length}
\label{sec:corrleng}

%
\begin{table*}[t]
  \begin{center}
  \begin{tabular}{cccccccc}
    \hline\hline & & & & & & &\\[-2ex]
    &Method& $\beta$ & $\delta$ & $\gamma$ & $\nu$ & $\nu_c$ & $\eta$ \\[1ex]
    \hline & & & & & & & \\[-2ex]
    $O(4)$ QM LPA (this work) & fRG Chebyshev &0.3989(41) &4.975(57) & 1.5458(68) &0.7878(25) &0.3982(17) &0 \\[1ex]
    $O(4)$ QM LPA$'$ (this work) & fRG Chebyshev &0.3832(31) &4.859(37) &1.4765(76) &0.7475(27)  & 0.4056(19)&0.0252(91)*\\[1ex]
    $Z(2)$ QM LPA (this work) & fRG Chebyshev &0.3352(12) &4.941(22) &1.3313(96)  &0.6635(17) &0.4007(45)&0\\[1ex]
    $Z(2)$ QM LPA$'$ (this work) & fRG Chebyshev &0.3259(01) &4.808(14) &1.2362(77) &0.6305(23) &0.4021(43)&0.0337(38)*\\[1ex]
    $O(4)$ scalar theories \cite{Tetradis:1993ts} &fRG Taylor&0.409& 4.80* & 1.556 & 0.791 &        & 0.034 \\[1ex]
    $O(4)$ KT phase transition \cite{VonGersdorff:2000kp} &fRG Taylor &0.387*& 4.73* &        & 0.739 &        & 0.047 \\[1ex]
    $Z(2)$ KT phase transition \cite{VonGersdorff:2000kp} &fRG Taylor &        &         &        & 0.6307  &        & 0.0467         \\[1ex]
    $O(4)$ scalar theories \cite{Litim:2001hk} & fRG Taylor & 0.4022*& 5.00* &    & 0.8043 &     &      \\[1ex]
    $O(4)$ scalar theories LPA\cite{Braun:2007td} &fRG Taylor&0.4030(30)& 4.973(30) &        & 0.8053(60) &       &        \\[1ex]    
    $O(4)$ QM LPA \cite{Stokic:2009uv} & fRG Taylor &  0.402 & 4.818 &  1.575 &  0.787  & 0.396 &  \\[1ex]
    $O(4)$ scalar theories \cite{Bohr:2000gp} & fRG Grid & 0.40 & 4.79 &     & 0.78 &     & 0.037 \\[1ex]
    $Z(2)$ scalar theories \cite{Bohr:2000gp} & fRG Grid & 0.32 & 4.75 &     & 0.64 &     & 0.044 \\[1ex]
    $O(4)$ scalar theories \cite{DePolsi:2020pjk} & fRG DE $\mathcal{O}(\partial^4)$ &   &    &     & 0.7478(9) &     & 0.0360(12) \\[1ex]
    $Z(2)$ scalar theories \cite{Balog:2019rrg,DePolsi:2020pjk} & fRG DE $\mathcal{O}(\partial^6)$ &   &    &     & 0.63012(5) &     & 0.0361(3) \\[1ex]
    $O(4)$ CFTs \cite{Kos:2015mba} & conformal bootstrap &   &    &     & 0.7472(87) &     & 0.0378(32) \\[1ex]
    $Z(2)$ CFTs \cite{Kos:2014bka} & conformal bootstrap &   &    &     & 0.629971(4) &     & 0.0362978(20) \\[1ex]
    $O(4)$ spin model \cite{Kanaya:1994qe}  &Monte Carlo&0.3836(46) & 4.851(22) & 1.477(18) & 0.7479(90) & 0.4019(71)* & 0.025(24)*  \\[1ex]
    $Z(2)$ $d=3$ expansion \cite{ZinnJustin:1999bf} & summed perturbation&0.3258(14) &4.805(17)*  &1.2396(13) &0.6304(13) &0.4027(23) &0.0335(25)      \\[1ex]
    Mean Field  && 1/2 &3&1&1/2&1/3 &0         \\[1ex]
    \hline\hline
  \end{tabular}
  \caption{Critical exponents for the $O(4)$ and $Z(2)$ symmetry universality classes in 3-$d$ space, obtained in the quark-meson LEFT within the fRG approach with truncations LPA and LPA$'$, where the effective potential is expanded as a sum of Chebyshev polynomials. Our calculated results are also in comparison to relevant results from previous fRG calculations, e.g., scalar theories with the effective potential expanded in a Taylor series \cite{Tetradis:1993ts,VonGersdorff:2000kp,Litim:2001hk,Braun:2007td}, or discretized on a grid \cite{Bohr:2000gp}, the quark-meson (QM) low energy effective theory with LPA \cite{Stokic:2009uv}, derivative expansions (DE) up to orders of $\mathcal{O}(\partial^4)$ and $\mathcal{O}(\partial^6)$ \cite{Balog:2019rrg,DePolsi:2020pjk}. Moreover, results from other approaches, such as the conformal bootstrap for the 3-$d$ conformal field theories (CFTs) \cite{Kos:2014bka,Kos:2015mba}, Monte Carlo simulation \cite{Kanaya:1994qe}, and $d=3$ perturbation expansion \cite{ZinnJustin:1999bf}, as well as the mean-field values of exponents are also presented. Note that values with an asterisk are obtained with scaling laws in \Eq{eq:relationExpo}.} 
  \label{tab:exponent}
  \end{center}\vspace{-0.5cm}
\end{table*}
%

It is well known that the correlation length $\xi$, plays a pivotal role in the critical dynamics, since fluctuations of wavelength $\sim\xi$ are inevitably involved in the dynamics. As a system is approaching towards a second-order phase transition, the most relevant degrees of freedom are the long-wavelength modes of low energy, and the correlation length is divergent at the phase transition \cite{Landau:1980}.

The critical behavior of correlation length is described by the critical exponent $\nu$, as shown in \Eq{eq:defiGamNu}. In the symmetric phase $t>0$, it reads
\begin{align}
  \xi&\sim t^{-\nu}\,,\qquad\mathrm{with}\qquad h=0\,,\label{}
\end{align}
which illustrates the scaling relation between the correlation length and the reduced temperature. Moreover, one can also define another critical exponent $\nu_c$ related to the scaling relation between the correlation length and the reduced external field, to wit,
\begin{align}
  \xi&\sim h^{-\nu_c}\,,\qquad\mathrm{with}\qquad t=0\,.\label{}
\end{align}

In our setup in the quark-meson LEFT, cf. \sec{sec:QM}, the correlation length is proportional to the inverse of the renormalized $\sigma$-meson mass, viz.,
\begin{align}
  \xi&\sim \frac{1}{m_\sigma}\,,\label{}
\end{align}
where $m_\sigma$ is related to the dimensionless $k$-dependent sigma mass $\bar{m}_{\sigma,k}$ in \Eq{eq:massmeson} via the relation as follows
\begin{align}
  m_\sigma&= \bar{m}_{\sigma,k}(\sigma=\sigma_{_{\mathrm{EoM}}})k\,,\quad\mathrm{with}\quad k\rightarrow 0\,,\label{}
\end{align}
where the scale $k$ is chosen to be in the IR limit $k\rightarrow 0$, and the mass is calculated on the equation of motion of the order parameter field. In \Fig{fig:nucnu} we show the scale relation between the correlation length and the reduced external field strength, and that between the correlation length and the reduced temperature, respectively. In the same way, we adopt the two different truncations: LPA and LPA$'$. The phase transition points are also chosen to be at the locations of the red and blue crosses in the phase diagrams in \Fig{fig:phasedia} for the $O(4)$ and $Z(2)$ universality classes, respectively. By the use of the linear fitting to the calculated data, one obtains values of the critical exponent $\nu$ as follows
\begin{align}
  \nu^{^{O(4)}}_{_{\mathrm{LPA}}}&=0.7878(25)\,,\qquad \nu^{^{O(4)}}_{_{\mathrm{LPA}'}}=0.7475(27)\,,\label{}\\[2ex]
  \nu^{^{Z(2)}}_{_{\mathrm{LPA}}}&=0.6635(17)\,,\qquad \nu^{^{Z(2)}}_{_{\mathrm{LPA}'}}=0.6305(23)\,,\label{}
\end{align}
as well as those of the critical exponent $\nu_c$, i.e.,
\begin{align}
  {\nu_c}^{^{O(4)}}_{_{\mathrm{LPA}}}&=0.3982(17)\,,\qquad {\nu_c}^{^{O(4)}}_{_{\mathrm{LPA}'}}=0.4056(19)\,,\label{}\\[2ex]
  {\nu_c}^{^{Z(2)}}_{_{\mathrm{LPA}}}&=0.4007(45)\,,\qquad {\nu_c}^{^{Z(2)}}_{_{\mathrm{LPA}'}}=0.4021(43)\,.\label{}
\end{align}

Finally, we close \sec{sec:exponent} with a summary of various critical exponents calculated in this work in \tab{tab:exponent}. Respective results for the $O(4)$ and $Z(2)$ symmetry universality classes with truncation LPA or LPA$'$ are presented in the first several rows in \tab{tab:exponent}. As we have discussed in \sec{sec:QM}, the effective potential is expanded as a sum of Chebyshev polynomials in our calculations, which captures global properties of the order-parameter potential very well. In \tab{tab:exponent} we also present values of critical exponents obtained from other computations, e.g., scalar theories calculated within the fRG with the effective potential expanded in a Taylor series \cite{Tetradis:1993ts,VonGersdorff:2000kp,Litim:2001hk,Braun:2007td}, or discretized on a grid \cite{Bohr:2000gp}, quark-meson LEFT within the fRG in LPA \cite{Stokic:2009uv}, derivative expansion of the fRG up to orders of $\mathcal{O}(\partial^4)$ and $\mathcal{O}(\partial^6)$ \cite{Balog:2019rrg,DePolsi:2020pjk}, the conformal bootstrap for the 3-$d$ conformal field theories \cite{Kos:2014bka,Kos:2015mba}, Monte Carlo simulation \cite{Kanaya:1994qe}, and the $d=3$ perturbation expansion \cite{ZinnJustin:1999bf}. One observes that our calculated results are in good agreement with the relevant results from previous fRG calculations as well as those from the conformal bootstrap, Monte Carlo simulation, and the $d=3$ perturbation expansion. Remarkably, the calculation with the truncation LPA$'$ is superior to that with LPA, and the former has already provided us with quantitative reliability for the prediction of the critical exponents in comparison to other approaches.

\section{summary}
\label{sec:summary}

QCD phase structure and related critical behaviors have been studied in the two-flavor quark-meson low energy effective theory within the fRG approach in this work. More specifically, we have expanded the effective potential as a sum of Chebyshev polynomials to solve its flow equation. Consequently, both the global properties of the effective potential and the numerical accuracy necessary for the computation of critical exponents are retained in our calculations. Moreover, we have employed two different truncations for the effective action: one is the usually used local potential approximation and the other is that beyond the local potential approximation, in which a field-dependent mesonic wave function renormalization is encoded.

With the numerical setup within the fRG approach described above, we have obtained the phase diagram in the plane of $T$ and $\mu_B$ for the two-flavor quark-meson LEFT in the chiral limit, including the second-order phase transition line of $O(4)$, the tricritical point and the first-order phase transition line. Furthermore, we also show the $Z(2)$ line in the phase diagram, which is the trajectory of the critical end point moving with the successive variance of the strength of explicit chiral symmetry breaking, or the varying pion mass.

In the phase diagram, we have performed detailed scaling analyses for the 3-$d$ $O(4)$ and $Z(2)$ symmetry universality classes, and investigated the critical behaviors in the vicinity of phase transition both in the chiral symmetric and broken phases. Moreover, the transverse and longitudinal susceptibilities of the order parameter have been calculated in the chiral broken phase near the coexistence line. 

A variety of critical exponents related to the order parameter, chiral susceptibilities and correlation lengths have been calculated for the 3-$d$ $O(4)$ and $Z(2)$ symmetry universality classes in the phase diagram, respectively. The calculated results are also compared with those from previous fRG calculations, either employing the Taylor expansion for the order-parameter potential or discretizing it on a grid, derivative expansion of the effective action, the conformal bootstrap, Monte Carlo simulations, and the $d=3$ perturbation expansion. We find that the critical exponents obtained in the quark-meson LEFT within the fRG approach, where the order-parameter potential is expanded in terms of Chebyshev polynomials and a field-dependent mesonic wave function renormalization is taken into account, are in quantitative agreement with results from approaches aforementioned. Furthermore, we have also investigated the size of the critical regime, and it is found that the critical region in the QCD phase diagram is probably very small, and it is smaller than 1 MeV in the direction of temperature.

\begin{acknowledgments}

We thank Jan M. Pawlowski for illuminating discussions. We also would like to thank other members of the fQCD collaboration \cite{fQCD} for work on related subjects. The work was supported by the National Natural Science Foundation of China under Contract No. 11775041, and the Fundamental Research Funds for the Central Universities under Contract No. DUT20GJ212.

\end{acknowledgments}

\appendix
\section{Threshold functions and anomalous dimensions}
\label{app:threshold}
We employ the $3d$ flat regulators \cite{Litim:2001up,Litim:2000ci} for quarks and mesons in this paper

\begin{align}
  R_{\phi,k}(q_0,\bm{q})&=Z_{\phi,k}\bm{q}^2 r_B(\bm{q}^2/k^2)\,, \label{eq:Rphi}\\[2ex] 
  R_{q,k}(q_0,\bm{q})&=Z_{q,k}i\bm{\gamma} \cdot \bm{q} r_F(\bm{q}^2/k^2)\,, \label{eq:Rq}
\end{align} 
with 
\begin{align}
  r_B(x)&=\left( \frac{1}{x}-1 \right)\Theta(1-x)\,,\\[2ex] 
  r_F(x)&=\left( \frac{1}{\sqrt{x}}-1 \right)\Theta(1-x)\,.  \label{}
\end{align} 
The threshold functions in \Eq{eq:flowV} are given by
\begin{align}
  &l_0^{(B,d)}(\bar{m}^{2}_{\phi,k},\eta_{\phi,k};T)\nonumber\\[2ex]
  =&\frac{2}{d-1}\left( 1- \frac{\eta_{\phi,k}}{d+1}\right)\frac{1}{\sqrt{1+\bar{m}^{2}_{\phi,k}}}\nonumber\\[2ex]
  &\times\bigg(\frac{1}{2}+n_{B}(\bar{m}^{2}_{\phi,k};T)\bigg)\,,\label{eq:l0B}
\end{align} 
and
\begin{align}
  &l_0^{(F,d)}(\bar{m}^{2}_{q,k},\eta_{q,k};T,\mu)\nonumber\\[2ex]
  =&\frac{2}{d-1}\left( 1-\frac{\eta_{q,k}}{d} \right)\frac{1}{2\sqrt{1+\bar{m}^{2}_{q,k}}}\nonumber\\[2ex]
    &\times\Big(1-n_{F}(\bar{m}^{2}_{q,k};T,\mu)-n_{F}(\bar{m}^{2}_{q,k};T,-\mu)\Big)\,.\label{eq:l0F}
\end{align} 
with the bosonic and fermionic distribution functions reading 
\begin{align}
n_{B}(\bar{m}^{2}_{\phi,k};T)=&\frac{1}{\exp \bigg\{\frac{k}{T}  \sqrt{1+\bar{m}_{\phi, k}^{2}}\bigg\}-1}\,,\label{}
\end{align}
and 
\begin{align}
n_{F}(\bar{m}^{2}_{q,k};T,\mu)=&\frac{1}{\exp\bigg\{\frac{1}{T}\Big[k\sqrt{1+\bar{m}^{2}_{q,k}}-\mu\Big]\bigg\}+1}\,,\label{}
\end{align}
respectively.

The meson anomalous dimension in \Eq{eq:etaphi} is given by
\begin{align}
  \eta_{\phi,k}(\rho)&=\frac{1}{6\pi^2}\Bigg\{\frac{4}{k^2} \bar{\rho}(\bar{V}''_k(\bar{\rho}))^2\mathcal{BB}_{(2,2)}(\bar{m}^{2}_{\pi,k},\bar{m}^{2}_{\sigma,k};T)\nonumber\\[2ex]
&+N_c\bar{h}^{2}_{y,k}\bigg[\mathcal{F}_{(2)}(\bar{m}^{2}_{q,k};T,\mu)(2\eta_{q,k}-3)\nonumber\\[2ex]
&-4(\eta_{q,k}-2)\mathcal{F}_{(3)}(\bar{m}^2_{q,k};T,\mu)\bigg]\Bigg\}\,, \label{eq:etaphi2}  
\end{align} 
with
\begin{align}
  \bar h_{y,k}&=\frac{h_{y,k}}{Z_{q,k}(Z_{\phi,k})^{1/2}}\,.\label{eq:barhk}
\end{align}
Note that threshold functions $\mathcal{BB}_{(2,2)}$, $\mathcal{F}_{(2)}$ and $\mathcal{F}_{(3)}$ in \Eq{eq:etaphi2} can be found in e.g., \cite{Fu:2015naa,Yin:2019ebz}.

\section{Some relations for the Chebyshev polynomials}
\label{app:cheby} 

In this appendix we collect some relations for the Chebyshev polynomials, which are used in solving the flow equation for the effective potential in \Eq{eq:barVk}. The Chebyshev polynomial of order $n$ reads
\begin{align}
 T_n(x)&=\cos\big(n\arccos(x)\big)\,,\label{}
\end{align}
with nonnegative integers $n$'s and $x\in[-1,1]$. The explicit expressions for the Chebyshev polynomials could be obtained by the recursion relation as follows
\begin{align}
 T_{n+2}(x)&=2x T_{n+1}(x)-T_{n}(x)\,,\quad n\geq 0\,,\label{}
\end{align}
with $T_{0}(x)=1$ and $T_{1}(x)=x$. 

The $N+1$ zeros of $T_{N+1}(x)$ in the region $-1\leq x\leq 1$ are given by
\begin{align}
 x_k&=\cos\left(\frac{\pi(k+\frac{1}{2})}{N+1}\right)\,,\quad k=0,\,1,\,\cdots N\,.\label{eq:zeros}
\end{align}
A discrete orthogonality relation is fulfilled by the Chebyshev polynomials, to wit, 
\begin{align}
  \sum_{k=0}^{N}T_i(x_k)T_j(x_k)&=\left \{\begin{array}{l}
 0\qquad\qquad\qquad i\neq j\\[1ex]
 (N+1)/2\qquad i=j\neq 0\\[1ex]
 N+1\qquad\quad\;\;\; i=j=0
\end{array} \right. \,, \label{eq:ortho}
\end{align}
where $x_k$'s are the $N+1$ zeros of $T_{N+1}(x)$ in \Eq{eq:zeros}, and $i,\,j\leq N$. The interval $[-1,1]$ for $x$ could be extended to an arbitrary one $[y_{\mathrm{min}},y_{\mathrm{max}}]$ for $y$ via the linear relation as follows
\begin{align}
 x&=\frac{2y-(y_{\mathrm{max}}+y_{\mathrm{min}})}{y_{\mathrm{max}}-y_{\mathrm{min}}}\,,\label{}
\end{align}
and the generalized Chebyshev polynomials are defined by
\begin{align}
 T_n^{[y_{\mathrm{min}},y_{\mathrm{max}}]}(y)&\equiv T_n\big(x(y)\big)\,.\label{eq:geneCheby}
\end{align}
Therefore, the zeros in $y$ corresponding to \Eq{eq:zeros} read
\begin{align}
 y_k&=\frac{y_{\max}-y_{\min}}{2}\cos\left(\frac{\pi(k+\frac{1}{2})}{N+1}\right)+\frac{y_{\max}+y_{\min}}{2}\,,\label{eq:zeros2}
\end{align}
with $k=0,\,1,\,\cdots N$. Then, a function $f(y)$ with $y\in [y_{\mathrm{min}},y_{\mathrm{max}}]$ can be approximated as 
\begin{align}
 f(y)&\approx\left[\sum_{i=1}^{N} c_{i} T_{i}^{[y_{\mathrm{min}},y_{\mathrm{max}}]}(y)\right]+\frac{1}{2} c_{0}\,,\label{eq:chebexpan}
\end{align}
where the coefficients could be readily obtained by the use of the orthogonality relation in \Eq{eq:ortho}, which yields
\begin{align}
 c_i&=\frac{2}{N+1} \sum_{k=0}^{N}f(y_k)T_{i}^{[y_{\mathrm{min}},y_{\mathrm{max}}]}(y_k)\,,\label{}
\end{align}
with $i=0,\,1,\,\cdots N$.

With the Chebyshev approximation of the function $f(y)$ in \Eq{eq:chebexpan}, it is straightforward to obtain its derivative, viz.
\begin{align}
 f'(y)&\approx\sum_{i=1}^{N} c_{i} \frac{d}{dy}T_{i}^{[y_{\mathrm{min}},y_{\mathrm{max}}]}(y)\nonumber\\[2ex]
&=\left[\sum_{i=1}^{N} d_{i} T_{i}^{[y_{\mathrm{min}},y_{\mathrm{max}}]}(y)\right]+\frac{1}{2} d_{0}\,,\label{eq:chebyDeri}
\end{align}
where the coefficients $d_{i}$'s can be deduced by the recursion relation, that reads
\begin{align}
 d_{N}&=0\,,\qquad d_{N-1}=\frac{2}{y_{\mathrm{max}}-y_{\mathrm{min}}}2 N c_{N}\,,\nonumber\\[2ex]
 d_{i-1}&=d_{i+1}+\frac{2}{y_{\mathrm{max}}-y_{\mathrm{min}}} 2i c_{i}\quad (i=N-1,\cdots,1)\,.\label{}
\end{align}
%


\bibliography{ref-lib.bib}

\end{document}